\documentclass[a4paper,12pt]{article}

\usepackage{amsmath}
\usepackage{color}
\usepackage{amssymb}
\usepackage{bm}
\usepackage{graphicx}

\begin{document}

\thispagestyle{empty}
\vspace*{-15mm}
{\bf OUJ-FTC-2}\\
{\bf OCHA-PP-352}\\


\vspace{8mm}
\begin{center}
{\Large\bf
Gravity Analog Model of Non-equilibrium Thermodynamics \\
}
\vspace{7mm}

\baselineskip 10pt
{\bf (The OUJ Tokyo Bunkyo Field Theory Collaboration)}

\vspace{2mm}
Noriaki Aibara${}^{1}$, Naoaki Fujimoto${}^{2}$, So Katagiri${}^{3}$,  Mayumi Saitou${}^{5}$, Akio Sugamoto${}^{4, 5}$, Takashi Yamamoto${}^{3}$ and Tsukasa Yumibayashi${}^{6}$

\vspace{2mm}

{\it ${}^{1}$Nature and Environment, Faculty of Liberal Arts, The Open University of Japan, Chiba 261-8586, Japan\\
${}^{2}$Department of Information Design, Faculty of Art and Design, Tama Art University, Hachioji, 192-0394 Japan \\
${}^{3}$Division of Arts and Sciences, The School of Graduate Studies,
The Open University of Japan, Chiba 261-8586, Japan\\
${}^{4}$Tokyo Bunkyo Study Center, The Open University of Japan (OUJ), \\
Tokyo 112-0012, Japan \\
${}^{5}$Ochanomizu University, 2-1-1 Ohtsuka, Bunkyo-ku, Tokyo 112-8610, Japan\\
${}^{6}$Department of Social Information Studies, Otsuma Women's University, 12 Sanban-cho, Chiyoda-ku, Tokyo 102-8357, Japan}

\end{center}

\vspace{5mm}
\begin{center}
\baselineskip 10pt
\noindent
\begin{abstract}
Non-equilibrium thermodynamics of Onsager and Machlup and of Hashitsume is reformulated as a gravity analog model, in which thermodynamic variables, kinetic coefficients and generalized forces form, respectively, coordinates, metric tensor and vector fields in a space of thermodynamic variables.  The relevant symmetry of the model is the general coordinate transformation.

Then, the entropy production is classified into three categories, when a closed path is depicted as a thermodynamic cycle.  One category is time reversal odd, and is attributed to the number of lines of magnetic flux  passing through the closed path, having monopole as a source.  There are two time reversal even categories, one of which is attributed to the space curvature around the path, having gravitational instanton as a source, which dominates for a rapid operation of the cycle.  The last category is the usual one, which remains even for the quasi-equilibrium operation.

It is possible to extend the model to include non-linear responses. In introducing new terms, important is the dimensional counting, using two parameters, the temperature and the relaxation time.   The effective action, being induced by the non-equilibrium thermodynamics, is derived.  This is a candidate for the action which controls the dynamics of kinetic coefficients and thermodynamic forces. An example is given in a chemical oscillatory reaction in a solvent of the van der Waals type.   Fluctuation-dissipation theorem is examined {\it \`a la} Onsager, and a derivation of the gravity analog thermodynamic model from quantum mechanics is sketched, based on an analogy to the resonance problem.
\end{abstract}

\end{center}

\newpage

\section{Introduction}

It is well known that the thermodynamics resembles the classical mechanics; an equilibrium state in the thermodynamics and an orbit of the classical mechanics are both given by a stationary point at which the entropy is maximum, and at which the action is extremum, respectively.  The essential difference exists between them, however, that the thermodynamics describes the ``dissipation'' while the classical mechanics describes the ``oscillation''.

The classical mechanics becomes the quantum mechanics at microscopic perspective, by including quantum fluctuations; the quantum mechanics is controlled by the probability, or the transition probability.  Similarly, if we include thermal fluctuations into the thermodynamics, temporal
development of the system is controlled by a transition probability $\Psi(\alpha,t | \alpha^{(0)}, t_0)$ from the state $\alpha_0$ at $t_0$ to the state $\alpha$ at $t$:
\begin{eqnarray}
\Psi(\alpha,t | \alpha^{(0)}, t_0) \propto \int  \mathcal{D} \alpha(t)~e^{-\frac{1}{2k_BT} \int_{t_0}^t dt'~\mathcal{L}^{\star}(\alpha(t'), \dot{\alpha}(t'))},
\end{eqnarray}
where, according to Onsager and Machlup\cite{Onsager-Machlup} or Hashitsume\cite{Hashitsume}, $\mathcal{L}^{\star}$ is given, in terms of thermodynamic variables $\alpha=(\alpha_1, \alpha_2, \cdots)$ (macroscopic variables), by
\begin{eqnarray}
\mathcal{L}^{\star}(\alpha, \dot{\alpha})=\Phi(\dot{\alpha}, \dot{\alpha})+\Phi^{(-1)} (X(\alpha), X(\alpha))- \sum_i \dot{\alpha}_i X_i(\alpha). \label{thermodynamic Lagrangian}
\end{eqnarray}
Here, $\Phi$ is the Lord Rayleigh's dissipation function,
\begin{eqnarray}
\Phi(\dot{\alpha}, \dot{\alpha})=\frac{1}{2} \sum_{i, j} R_{ij} \dot{\alpha}_i \dot{\alpha}_j, ~~\Phi^{(-1)}(X, X)=\frac{1}{2} \sum_{i, j} L_{ij} X_i X_j,
\end{eqnarray}
and $\dot{\alpha}$ (time derivative of $\alpha$) is a current driven by thermodynamic forces $X_i(\alpha)$; they satisfy the ``constitutional equation'' in case without thermal fluctuations at zero temperature:
\begin{eqnarray}
\dot{\alpha}_i=\sum_{j} L_{ij} X(\alpha)_j,~\mathrm{at}~T=0,
\end{eqnarray}
where the kinetic coefficients are denoted by $L_{ij}$; $R_{ij}$ (resistance) is the  inverse matrix of $L_{ij}$ (conductivity), satisfying $\sum_k R_{ik}L_{kj}=\delta_{ij}$.  This is a linear response theory, since the current $\dot{\alpha}$ is a linear function of the force $X$.  See for example, Eq. (9.20) in the essay in \cite{Hashitsume}.
Hereafter, we put ``star" mark on the ``Lagrangian'' in thermodynamics, since it describes the ``dissipation" and is completely foreign to the usual Lagrangian in classical mechanics and in quantum mechanics which describes the ``oscillation''.

If we treat the thermodynamic variables $\alpha=(\alpha_1, \alpha_1, \dots)$ as coordinates $x^{\mu}=(x^1, x^2, \dots)$, then a space $M$, spanned by these coordinates, forms a manifold of $n$ dimensions, which we will call ``Thermodynamic Space (ThS)''.\footnote{Time $t$ is not included in $x^{\mu}=(x^1, x^2, \dots)$. Depending on the problem, $n$ is properly chosen.  For a chemical reaction in a solution, $n$ can be the number of species of the solute.  If the number of molecules of $i$-th solute is $N_i$ and its chemical potential is $\mu_i$, then $(x^1, \cdots, x^n)=(N_1, \cdots, N_n)$ and $(A_1, \cdots, A_n)=(\mu_1, \cdots, \mu_n)$.  We can study also the heat and electric conductions. In such cases, $n>>1$.  For the heat flow as an example (see \cite{Hashitsume}), we have to consider the spacial dependency of temperature; we use $\{x_1, x_2, \cdots, x_n\}$ as a ``label of each portion'' with equal mass of a given substance. The thermodynamic variable is the temperature at different portions, $X(x, t)\equiv(T(x, t)-T^0)/T^0$, and hence $n>>1$.  When a three-dimensional substance is divided into $N$ portions, then $n=3N$. In this case, the constitutional equation before fluctuations are introduced is $\dot{X}(x, t)=\int d^n y~g(x, y) A(y)$, where the kinetic coefficient (heat conductivity) and the thermodynamic force are given by $g(x, y)=-\frac{\kappa T^0}{c^2} \partial_y^{2} \delta^{(n)}(x-y)$ and $A(y)=-\frac{c}{T^0} X(y) $, respectively. The heat capacity per portion is $c$, the heat conductivity in the Fourier law is $\kappa$, and $T^0$ is a constant temperature after thermal equilibrium is realized. }

Then, we can identify the thermodynamic forces $X_i(\alpha)$ as a vector field $A_{\mu}(x)$, and the kinetic coefficients $R_{ij}$ and its inverse $L_{ij}$ as a metric tensor $g_{\mu\nu}(x)$ and its inverse $g^{\mu\nu}(x)$ of the thermodynamic space $M$, that is we have the following correspondence:
\begin{eqnarray}
\alpha_i \leftrightarrow x^{\mu}, ~X_i(\alpha) \leftrightarrow A_{\mu}(x), L_{ij} \leftrightarrow g^{\mu\nu}(x),~\mathrm{and}~ R_{ij} \leftrightarrow g_{\mu\nu}(x).
\end{eqnarray}
Then, Eq.(\ref{thermodynamic Lagrangian}) becomes
\begin{eqnarray}
\mathcal{L}^{\star}_1(x, \dot{x})=\frac{1}{2} g_{\mu\nu}(x) \dot{x}^{\mu}\dot{x}^{\nu} + \frac{1}{2} g^{\mu\nu}(x) A_{\mu}(x) A_{\nu}(x) -\dot{x^{\mu}} A_{\mu}(x) \label{field theoretical Lagrangian 1},
\end{eqnarray}
and we arrive at a ``gravity analog model of thermodynamics'', which is a natural and an easily tractable model as gauge and gravitational theories.

We have to note that $x^{\mu}$ may have different dimensions for different $\mu$, since the thermodynamic variables have various dimensions.  In the same manner the gauge fields and the metric tensors have different dimensions for different components.  The relation, $\sum_{j} L_{ij}R_{jk}=\delta_{ik}$ between two kinetic coefficients $R_{ij}$ and $L_{ij}$ is naturally reproduced by $g_{\mu\nu}g^{\nu\lambda}=\delta_{\mu}^{\lambda}$.
At this point we are tempted to introduce the kinetic terms of gauge fields and of metric, such as
\begin{eqnarray}
\mathcal{L}^{\star}_2(A_{\mu}, g_{\mu\nu})=-\frac{1}{4} C_1 g^{\mu\lambda}(x)g^{\nu\rho}(x) F_{\mu\nu}(x) F_{\lambda\rho}(x) + C_2 R(g_{\mu\nu}(x)), \label{field theoretical Lagrangian 2}
\end{eqnarray}
where $F_{\mu\nu}(x)=\partial_{\mu}A_{\nu}(x)- \partial_{\nu}A_{\mu}(x)$ is the field strength, $g(x)=\mathrm{det}g_{\mu\nu}(x)$, and $R(g_{\mu\nu}(x))$ is the scalar curvature of the metric $g_{\mu\nu}(x)$.  However, we will discuss this problem in the later sections when the proper symmetry of the system will be manifest.

There are a number of works in which the thermodynamics is constructed as a Riemann geometry. For example in  \cite{Ruppeimer}, the thermodynamic force and the metric are defined using the entropy $S(x)$ by $A_{\mu}(x)=T \partial S/\partial x^{\mu}$ and $g_{\mu\nu}(x)=\partial^2 S/\partial x^{\mu} \partial x^{\nu}$, respectively; then, the scalar curvature of thermodynamically constructed Riemann manifold can be connected to the correlation length of phase transition dynamics at critical point.  In the final stage of writing this paper we have noticed the paper by Sonnino and Sonnino in \cite{Sonnino}, which studied the physical meaning of general coordinate invariance in thermodynamics (they called Thermodynamic Covariance Principle) based on the discussion of entropy production and the Glansdorff-Prigogine dissipative quantity. Their gravity analog model is more radical than ours, since the manifold is not Riemannian, but includes an antisymmetric part of the metric tensor. Our paper does not overlap so much with their work, except for the identification of the symmetry of thermodynamics to the general covariance.

In the next section, a derivation of the thermal Lagrangian is given by introducing thermal fluctuations in the constitutional equation {\it \`a la} Langevin equation.  In Sec. 3, the symmetry of the gravity analog thermodynamic model is clarified, and the dimensionality of various quantities is controlled by using two constants, one is the temperature $T$, and the other is a relaxation time $t_{\star}$. In Sec. 4, possible extension of the model is discussed; the extension to non-linear responses and the inclusion of kinetic terms for metric $g_{\mu\nu}(x)$ and for vector (gauge) field $A_{\mu}(x)$. In Sec. 5, entropy production in the thermal processes is studied and categorized.  In Sec. 6, the effective action induced by the non-equilibrium thermodynamics is derived.   In the subsequent section (Sec. 7), the source of the entropy production is attributed to the magnetic flux passing through a closed path depicted by a thermal process, or to the space curvature around the path.  In Sec. 8, an example of the gravity analog model is given, in a chemical reaction in a solvent.  The fluctuation-dissipation theorem is examined in Sec.9 .  In Sec. 10 , a derivation of the gravity analog thermodynamic model from quantum mechanics is sketched, in which we clarify that the role of Hamiltonian operator in quantum mechanics is played by the thermodynamic operator which describes the decay rate (decay width) in thermodynamics.

The last section is devoted to conclusion and discussions.  Appendix A is prepared for the preliminaries of the example. In Appendix B,  Fokker-Planck equation and the classical description of thermodynamics are examined, from which the operator formalism of thermodynamics can be extracted.
\section{Derivation of the thermodynamic Lagrangian}
In order to understand the dynamics of the gravity analog thermodynamic model given in the last section, we first derive its Lagrangian Eq.(\ref{field theoretical Lagrangian 1}) from the Langevin equation.

If we incorporate thermal fluctuations, the constitutional equation is modified by random or stochastic forces $\xi^{\mu}(x, t)$ as follows:
\begin{eqnarray}
\dot{x}^{\mu}-g^{\mu\nu}(x)A_{\nu}(x) = \xi^{\mu}(x, t), \label{x and xi, 1}
\end{eqnarray}
which can be understood as follows: The thermal force has two parts, one is a macroscopic force being able to specify by $g^{\mu\nu}(x)A_{\nu}(x)\equiv A^{\mu}(x)$, the other is a microscopic random force $\xi^{\mu}(x, t)$, being unable to specify; we can only assume the strength squared of this random force is, on average, proportional to the temperature $T$. This kind of random force follows the Gaussian probability distribution $Pr[\xi]$,
\begin{eqnarray}
Pr[\xi] \propto e^{-\frac{1}{2k_B T} \int_{-\infty}^{+\infty} dt~\frac{1}{2}g_{\mu\nu}(x) \xi^{\mu}(x, t) \xi^{\nu}(x, t)} \equiv e^{-\frac{1}{2k_B T} \int_{-\infty}^{+\infty} dt~\frac{1}{2} \xi_{\mu}(x, t)\xi^{\mu}(x, t)},
\end{eqnarray}
since it gives
\begin{eqnarray}
\left<\left< \xi^{\mu}(x, t)\right>\right>_{\xi} = 0, ~~\left<\left< \xi^{\mu}(x, t)\xi^{\nu}(x, s) \right>\right>_{\xi} = 2 k_B T~ g^{\mu\nu}(x)~ \delta(t-s),
\end{eqnarray}
where $\left<\left< \dots \right>\right>_{\xi}$ denotes the Gaussian average over $\xi$ by $Pr[\xi]$. Explicitly, it is given by
\begin{eqnarray}
\left<\left< O(\xi) \right>\right>_{\xi} = \frac{\int \prod_{\mu} \mathcal{D} \xi^{\mu}(x, t) \sqrt{g(x)}~ O(\xi)~ e^{-\frac{1}{2k_B T} \int_{-\infty}^{+\infty} dt~\frac{1}{2}g_{\mu\nu}(x) \xi^{\mu}(x, t) \xi^{\nu}(x, t)}}{\int \prod_{\mu} \mathcal{D} \xi^{\mu}(x, t) \sqrt{g(x)} ~e^{-\frac{1}{2k_B T} \int_{-\infty}^{+\infty} dt~\frac{1}{2}g_{\mu\nu}(x) \xi^{\mu}(x, t) \xi^{\nu}(x, t)}},
\end{eqnarray}
where $g(x)=\det g_{\mu\nu}(x)$.

The random variable $\xi^{\mu}$ and the thermodynamic variable $x^{\mu}$ are connected by Eq.(\ref{x and xi, 1}).  Therefore, an operator $\hat{O}(t_1, t_2, \dots t_N)_x$ given in terms of thermodynamic variables $x$, can be expressed as $\hat{O}(t_1, t_2, \dots t_N)_{\xi}$ in terms of the random variable $\xi$.  Starting from the expectation value averaged over the random variable, that is expressed in terms of the path integral over the random variables, we can obtain the expectation value over the thermodynamic variables, expressed in terms of the path integration over $x^{\mu}(t)$.\footnote{The correspondence between a path in $\xi$-space, $(\xi_0, \xi_1, \cdots, \xi_n)$, and a path in $x$-space, $(x_0, x_1, \cdots, x_n)$, is given by $\xi^{\mu}_i=\xi^{\mu}(x^{\nu}_i, t_i)$, where $t_0 < t_1 < \cdots < t_{n}=t$ is the discretized time sequence.}
\begin{eqnarray}
\left<\left<  \hat{O}(t_1, t_2, \dots t_N)_{\xi} \right>\right>_{\xi} = \left< \hat{O}(t_1, t_2, \dots t_N)_{x} \right>_{x},
\end{eqnarray}
where we add the suffices $\xi$ and $x$ to differentiate the two different expectation values.  If we pay attention to the Jacobian which does appear during the change of variables from $\xi$ to $x$, we have the following result:
\begin{eqnarray}
\left< \hat{O}(t_1, t_2, \dots t_N)_{x} \right>_{x} \propto \int \prod_{\mu} \mathcal{D}x^{\mu}(t)  \sqrt{g(x)}~\hat{O}(t_1, t_2, \dots t_N)_{x} ~e^{-\frac{1}{2k_b T} \int dt \mathcal{L}^{\star}_3(t)},
\end{eqnarray}
where $\mathcal{L}^{\star}_3$ is given by
\begin{eqnarray}
\mathcal{L}^{\star}_3= \mathcal{L}^{\star}_1 + \mathcal{L}^{\star}_{\mathrm{ghost}}.
\end{eqnarray}
Here $\mathcal{L}^{\star}_{\mathrm{ghost}}$ is given by
\begin{eqnarray}
\mathcal{L}^{\star}_{\mathrm{ghost}}=\bar{c}_{\mu}(t) \frac{\partial \xi^{\mu}}{\partial x^{\nu}} c^{\nu}(t)=\bar{c}_{\mu}(t) \left[ \delta^{\mu}_{\nu} \partial_t - \partial_{\nu}(g^{\mu\lambda}(x) A_{\lambda}(x) ) \right] c^{\nu}(t),
\end{eqnarray}
where $c^{\nu}$ and $\bar{c}_{\mu}$ are fermionic ghost and anti-ghost particles, the effect of them is only to reproduce Jacobian $\det \left[\partial \xi^{\mu}/\partial x^{\nu}\right]$ which appears in the change of variables.
This is what Parisi and Sourlas did in 1979 \cite{Parisi-Sourlas}.  This Lagrangian can be written in a form familiar to the gauge fixed Lagrangian.  That is,
\begin{eqnarray}
\mathcal{L}^{\star}_4=-\frac{1}{2} g^{\mu\nu}(x) b_{\mu}(t) b_{\nu}(t) + b_{\mu}(t) \xi^{\mu}(t)_x + \bar{c}_{\mu}(t) \frac{\partial \xi^{\mu}_x }{\partial x^{\nu}}c^{\nu}(t),
\end{eqnarray}
where $\xi_x$ is a function of $x$.  The $\mathcal{L}^{\star}_3$ and $\mathcal{L}^{\star}_4$ are identical, as is shown by the path integration over the Lagrange multiplier field $b_{\mu}$ (sometimes called Nakanishi-Lautrup field).

Now we discuss a meaning of the dynamics using $\mathcal{L}^{\star}_4$. When a local gauge symmetry $\delta x^{\mu}= \epsilon^{\mu}(t) $ (a local translation), is fixed, the symmetry reduces to the global supersymmetry transformation by $s$ {\it \`a la} BRST \cite{BRST}, where the original gauge transformation parameter $\epsilon^{\mu}(t)$ is replaced by a ghost particle $c^{\mu}(t)$ \cite{Baulieu}:
\begin{eqnarray}
s x^{\mu}= c^{\mu}, ~sc^{\mu}=0, ~s\bar{c}_{\mu}=b_{\mu}, ~sb_{\mu}=0.
\end{eqnarray}

This means the Lagrangian before gauge fixing is $\mathcal{L}^{\star}_0=0$, a trivial theory or a ``topological theory'' without any dynamics.  It has the infinitesimal gauge symmetry of $\delta x^{\mu}= \epsilon^{\mu}$, the symmetry of infinitesimal variation of thermodynamic variables.  After gauge fixing by the Feymann gauge with a gauge fixing function $\xi_x$ and the gauge parameter $\alpha=1$, the dynamics appears, which reproduces the thermally non-equilibrium thermodynamics of Onsager and Machlup \cite{Onsager-Machlup} and of Hashitsume \cite{Hashitsume}.\footnote{If we take the Landau gauge with $\alpha=0$, then the thermal fluctuations disappear.}

Recently one of the authors (So Katagiri) has considered the gauge symmetry and its gauge fixing more seriously, and arrived at an interesting understanding of thermodynamics as a gauge theory with its fixing\cite{Katagiri}.

Without introducing ghost fields and supersymmetry, the model can be written as follows:
\begin{eqnarray}
\mathcal{L}^{\star}_1&=&\frac{1}{2} g_{\mu\nu}(x) \dot{x}^{\mu}\dot{x}^{\nu} + \frac{1}{2} g^{\mu\nu}(x) A_{\mu}(x) A_{\nu}(x) -\dot{x}^{\mu} A_{\mu}(x) +2k_BT \sum_{n=1}^{\infty} \frac{1}{n} \idotsint_{-\infty}^{\infty} dt_1 \cdots dt_n  \nonumber \\
&\times& Tr \left( \theta(t_n-t_1) \hat{M}(t_1)\theta(t_1-t_2)\hat{M}(t_2) \cdots \theta(t_{n-1}-t_n)\hat{M}(t_n)\right), \label{Lagrangian star 1}
\end{eqnarray}
where the matrix is defined by $(\hat{M}(t))^{\mu}_{~~\nu}=\partial_{\nu} A^{\mu}(x(t))$, and the step  function $\theta(t-s)$ is the ghost propagator satisfying $\partial_t \theta(t-s)=\delta(t-s)$.\footnote{To show this, $\det (1-M) = e^{\mathrm{Tr} \ln (1-M)}=e^{-\mathrm{Tr} \sum_n M^n/n}$ is used.}

This section is written, using the standard terminology of Langevin equation and thermal fluctuation theory in thermodynamics, but for many people, it is easier to consider $\xi_{\mu}$ as the momentum $p_{\mu}$ which is  ``thermodynamically conjugate'' to $x^{\mu}$.  This viewpoint may help them to understand the structure of thermodynamics more clearly.  See Appendix for this purpose.

\section{Symmetry of the thermodynamic model}

We have to elucidate what is the proper symmetry of our gravity analog model of thermodynamics.

It is recognized that the usual gauge symmetry for a vector field, $A_{\mu} \to A_{\mu} + \partial_{\mu} \epsilon(x)$, is broken from the beginning by the existence of $A_{\mu}A^{\mu}$ term which is inevitable to represent the thermodynamic force $\sum_{i, j} L_{ij}X_iX_j$.  The symmetry breaking term is a mass term of gauge field, so that it is interesting to consider this term as evidence of the spontaneously broken symmetry.   In this paper, however, we do not adopt the usual gauge symmetry as a symmetry of the thermodynamics.  On the other hand, the general coordinate invariance (or diffeomorphism invariance) seems to be a proper symmetry, so that we adopt it as the symmetry of thermodynamics.  It is natural to impose the concept of relativity in thermodynamics; the thermodynamical system be relatively equivalent between two different frames using different thermodynamic variables; the variables in two frames are connected naturally by the general coordinate transformation:
\begin{eqnarray}
x^{\mu} \to x^{'\mu}=f^{\mu}(x),
\end{eqnarray}
where $f^{\mu}(x)$ can be any  (differentiable) function.

Now the vector field $A_{\mu}$ representing the thermodynamic force becomes a covariant vector, satisfying
\begin{eqnarray}
A_{\mu}(x) \to A'_{\mu}(x')=\frac{\partial x^{\nu}}{\partial x^{'\mu}} A_{\nu}(x),
\end{eqnarray}
$A^{\mu}(x)$ is a contravariant vector, satisfying
\begin{eqnarray}
A^{\mu}(x) \to A^{'\mu}(x')=\frac{\partial x^{'\mu}}{\partial x^{\nu} } A^{\nu}(x),
\end{eqnarray}
and the metric (or the set of kinetic coefficients or resistances) becomes covariant tensor,
\begin{eqnarray}
g_{\mu\nu}(x) \to g'_{\mu' \nu'}(x')= \frac{\partial x^{\mu}}{\partial x^{'\mu'}}\frac{\partial x^{\nu}}{\partial x^{'\nu'}}   g_{\mu\nu}(x).
\end{eqnarray}

Under this general coordinate transformation, or under the general transformation of thermodynamic variables (or diffeomorphism), the Lagrangian $\mathcal{L}^{\star}_1$ is invariant, including the last Jacobian terms.\footnote{Before eliminating the microscopic random force, $\xi^{\mu}(x, t)$ is a contravariant vector and $\xi_{\mu}(x, t)$ is a covariant vector. This is consistent with Eq.(\ref{x and xi, 1}).}

Therefore, we will consider in this paper, ``the general coordinate transformation'' (the general transformation of thermodynamic variables) is the proper symmetry of our thermodynamic model $\mathcal{L^{\star}}$. We hope to describe the irreversible process and the thermal non-equilibrium state in the model.

Here, we stop for a while and do the ``dimensional analysis'' of various variables, since thermodynamic variables have various dimensionalities.  We use the notation $[O]$ to represent the ``dimension'' of a variable $O$.  The entropy has the same dimension as the Boltzmann constant, namely $[S]=[k_B]$.  Therefore, the thermodynamic force $X_i=T\partial S/\partial \alpha_i$ (even if it is a classical expression, the dimensionality does not change after including fluctuations.) has the dimension $[X_i]=[E_{\star}]/[\alpha_i]$, where $E_{\star}$ is a basic energy of the system.  From the constitutional equation of $\dot{\alpha_i}-L_{ij} X_j=\xi_i$, we have $[L_{ij}]=[\alpha_i][\alpha_j]/[E_{\star}][t_{\star}], [R_{ij}]=[E_{\star}][t_{\star}]/[\alpha_i][\alpha_j]$, and $[\xi_i]=[\alpha_i]/[t_{\star}]$, where $t_{\star}$ is a basic time scale of the system.  The dimension of $\alpha_i$ can be freely chosen.

In terms of $(x^{\mu}, A_{\mu}, g_{\mu\nu})$ we have
\begin{eqnarray}
[A_{\mu}]=\frac{[E_{\star}]}{[x^{\mu}]}, ~[\partial_{\mu}]=\frac{1}{[x^{\mu}]}, ~[g^{\mu\nu}]=\frac{[x^{\mu}][x^{\nu}]}{[E_{\star}][t_{\star}]}, ~[g_{\mu\nu}]=\frac{[E_{\star}][t_{\star}]}{[x^{\mu}][x^{\nu}]}, [\xi^{\mu}]=\frac{[x^{\mu}]}{[t_{\star}]}.
\end{eqnarray}
In the situation that all vector indices are contracted to form a scalar, the dimensions coming from $[x^{\mu}]$ cancel with themselves.  Therefore, only the dependence on $[E_{\star}]$ and $[t_{\star}]$ remains.  Ignoring the unnecessary dependence on $[x^{\mu}]$, we can consider the dimension as follows:
\begin{eqnarray}
[A_{\mu}]' = [E_{\star}], ~[A^{\mu}]' = \frac{1}{[t_{\star}]}, ~[\partial_{\mu}]'=1, ~[g^{\mu\nu}]'=\frac{1}{[E_{\star}][t_{\star}]}, ~[g_{\mu\nu}]'=[E_{\star}][t_{\star}], [\xi^{\mu}]'=\frac{1}{[t_{\star}]}, \label{effective dimension}
\end{eqnarray}
but it is not recommended to use $[x_{\mu}]'$ and $[\partial^{\mu}]'$ in the dimensional counting, since they have non-trivial values, $[x_{\mu}]'=[E_{\star}][t_{\star}]$ and $[\partial^{\mu}]'=1/([E_{\star}][t_{\star}])$, respectively.

Now, we can easily recognize, each term in $\mathcal{L}^{\star}_1$ in Eq.(\ref{Lagrangian star 1}) has the dimension of $[E_{\star}]/[t_{\star}]$, so that the action can be dimensionless as in $(2k_B T)^{-1} \int dt \mathcal{L}^{\star}_1$.  Here the basic energy $E_{\star}$ is chosen to be $k_B T$ using the temperature $T$ of the system.  It is possible to consider the temperature as a local scalar field having $x$ dependency. However, we consider $T$ to be constant in this paper.  The other basic constant of time $t_{\star}$ may be chosen to be the relaxation time $\tau$ for the system to attain equilibrium.  This also can be locally defined as $\tau(x)$, which gives the relaxation time in the neighborfood of the point $x$ in ThS.

\section{Possible extension of the model}
So far the model is nothing but the model of Onsager and Machlup \cite{Onsager-Machlup}.  We have changed the viewpoint and considered the space of the thermodynamic variables (ThS) as the manifold (probably curved manifold) which is familiar in the general relativity; we have considered a thermodynamic force as a vector field $A_{\mu}(x)$, and the transport coefficients as the metric tensor $g_{\mu\nu}(x)$ in the manifold.  Then, the proper symmetry of this thermodynamic space (ThS) will become manifestly the general coordinate invariance (general covariance) of the space.  The Lagrangian $\mathcal{L}^{\star}_1$ is general coordinate invariant; in the Lagrangian the Jacobian terms were not included in Onsager and Machlup\cite{Onsager-Machlup} and others, but they do exist as was pointed out by Parisi and Sourlas\cite{Parisi-Sourlas}.

Without violating the proper symmetry, {\it i.e.}, the general covariance, we can add any term of $A_{\mu}(x)$ and $g_{\mu\nu}(x)$ into the Lagrangian.

First we generalize the interaction between $x^{\mu}$ and $A_{\mu}$. This is related to the generalization of the constitutional equation. The simplest term that appeared in $\mathcal{L}^{\star}_1$ reproduces the ordinary linear response theory.  To extend it to non-linear response theories, the Langevin method is useful, since a simple extension of Eq.(\ref{x and xi, 1}) introduces the non-linearity.  Even restricting to the lowest extension, we have\footnote{In the following we always use the standard notation of derivatives $\partial_{\mu}$. To obtain the general coordinate invariant expression, however, these derivatives should be properly replaced by the covariant derivatives $\nabla_{\mu}$, including the effect of curved space.}
\begin{eqnarray}
\xi^{\mu}= \dot{x}^{\mu}-A^{\mu}(x) -c_1 A^{\mu}(x) (\partial_{\nu}A^{\nu}(x))
-c_2 \partial^{\mu} (A_{\nu}(x) A^{\nu}(x)) - c_3 A^{\mu}(x) A_{\nu}(x)A^{\nu}(x) - \cdots.
\end{eqnarray}
  In general, the terms such as $R^{\mu}_{~\nu} A^{\nu}$, $R A^{\mu}$, and others can appear in the right hand side of the above equation.
By the help of the dimensional analysis Eq.(\ref{effective dimension}) studied in the last section, we have
\begin{eqnarray}
\xi^{\mu}&=& \dot{x}^{\mu}-A^{\mu}(x) -\tau\left\{ c'_1A^{\mu}(x) (\partial_{\nu}A^{\nu}(x))
+c'_2 \partial^{\mu} (A_{\nu} (x)A^{\nu}(x)) +  \frac{c'_3}{k_B T} A^{\mu}(x) A_{\nu}(x)A^{\nu}(x)\right\} \nonumber \\
&&+ \cdots,
\end{eqnarray}
where the coefficients with prime are dimensionless numerical constants.

This can incorporate the non-linear response effects, giving the constitutional equation at $T=0$ as
\begin{eqnarray}
\dot{x}^{\mu}= A^{\mu}(x)+  \tau \left\{ c'_1 A^{\mu}(x) \partial_{\nu}A^{\nu}(x)
+c'_2 \partial^{\mu} (A_{\nu}(x) A^{\nu}(x))+\frac{c'_3}{k_B T} A^{\mu}(x) A_{\nu}(x)A^{\nu}(x)\right\} + \cdots.
\end{eqnarray}
In terms of the thermodynamic words, this is represented by
\begin{eqnarray}
\dot{\alpha}_{i}&=& \sum_j L_{ij} X_j(\alpha)+  \tau \sum_{jkl} L_{ij} L_{kl} \left\{ c'_1X_j(\alpha) \partial_{k}X_l(\alpha)+c'_2  \partial_{j} X_{l}(\alpha) X_{k}(\alpha)+\frac{c'_3}{k_B T} X_j(\alpha) X_k(\alpha) X_l(\alpha) \right\} \nonumber \\
&&+ \cdots.
\end{eqnarray}

This extension leads to the following $\mathcal{L}^{\star}_{1,\mathrm{NL} } $ via $\xi_{\mu}\xi^{\mu}$, that is,
\begin{eqnarray}
&&\mathcal{L}^{\star}_{1, NL}= \left[ \dot{x}^{\mu}-A^{\mu}(x) -\tau \left\{ c'_1 A^{\mu}(x) \partial_{\nu}A^{\nu}(x) + c'_2 \partial^{\mu} A_{\nu} (x)A^{\nu}(x) + \frac{c'_3}{k_B T}~A^{\mu}A_{\nu}A^{\nu}\right\} + \cdots \right] \nonumber \\
&&~\times \left[ \dot{x}_{\mu}-A_{\mu}(x) -\tau\left\{ c'_1 A_{\mu}(x) \partial_{\lambda}A^{\lambda} + c'_2 \partial_{\mu} A_{\lambda}(x) A^{\lambda}(x) + \frac{c'_3}{k_BT} A_{\mu}A_{\lambda}A^{\lambda}\right\} + \cdots \right].
\end{eqnarray}

Here we understand the importance of the dimensional counting of each term in the action, by using two parameters $T$ and $\tau$.  For example, let us compare the linear response and the non-linear response in the above example.  The non-linear term $A^{\mu} A_{\lambda} A^{\lambda}$ has an extra factor $\tau/k_BT$ relative to the linear response term $A^{\mu}$.  This means that if the energy scale $E$ and the time scale $t$ of the problem we are examining satisfy
\begin{eqnarray}
\frac{E}{k_BT} >> \frac{t}{\tau},
\end{eqnarray}
then the non-linear response dominates over the linear response, which occurs in the phenomena with a slow relaxation time $(\tau>>t)$ or at low temperature $(T << E)$.

In the same way as in the linear response case, the thermal fluctuations in the non-linear response case, can also be taken into account by
\begin{eqnarray}
\left<\left<  \hat{O}(x) \right>\right>_{\xi} =\left<  \hat{O}(x) \right>_{x} \propto \int \mathcal{D} x^{\mu}(t) \sqrt{g(x(t))} ~O(x) ~e^{-I^{\star}_1(x)},
\end{eqnarray}
with the action  $I^{\star}_1= \frac{1}{2k_BT}\int dt \mathcal{L}^{\star}_{1, NL}(x(t))$. Here $g(x)=\mathrm{det}g_{\mu\nu}(x)$ as before, and in doing the path integration,
$\int \mathcal{D} x^{\mu}(t) \sqrt{g(x(t))}=\prod_{t} \left(dx^1(t) \cdots dx^n(t)\sqrt{g(x(t))}\right)$ is the invariant combination under the general coordinate transformation.
So far the $A_{\mu}(x)$ and $g_{\mu\nu}(x)$ are background fields given in the thermodynamic space (ThS).  However, if the $A_{\mu}(x)$ and $g_{\mu\nu}(x)$ start to fluctuate thermally, like the quantum fluctuations in the gauge and the gravitational theories, then we have to path-integrate over $A_{\mu}(x)$ and $g_{\mu\nu}(x)$, with a proper probability.  Again the way to determine the probability is to impose the proper symmetry of the system, or to impose the general covariance.  The candidates are easily written down as follows:
\begin{eqnarray}
\mathcal{L}^{\star}_2&=&-\frac{1}{4} a_1 g^{\mu\lambda}(x)g^{\nu\rho}(x) F_{\mu\nu}(x) F_{\lambda\rho}(x) +a_2 g^{\mu\lambda}(x) g^{\nu\rho}(x) A_{\mu}(x) A_{\nu}(x) A_{\lambda}(x) A_{\rho}(x) \nonumber \\
&&-\frac{1}{2} a_3  \left(g^{\mu\nu}(x) \partial_{\mu} A_{\nu}\right)^2 + \cdots + b_1  R(x)  + b_2 R_{\mu\nu}(x) A^{\mu}(x)A^{\nu}(x)  + \cdots, \label{Lagrangian star 2}
\end{eqnarray}
where $F_{\mu\nu}(x)=\partial_{\mu}A_{\nu}(x)- \partial_{\nu}A_{\mu}(x)$, $R(x)$ is the scalar curvature, and $R_{\mu\nu}(x)$ is the Ricci tensor  for the metric $g_{\mu\nu}(x)$.

It is noted in Eq.(\ref{Lagrangian star 2}), in addition to the usual transverse part of $A_{\mu}$ as in the $a_1$ term, the longitudinal part $A_{\mu}$ appears in $a_3$  term, since the usual gauge symmetry of restricting $A_{\mu}$ to transverse modes is absent here.  In other words, the thermodynamics can be understood as a gauge theory of the longitudinal parts. As for the gravity part, there are a lot of other possibilities which appear in various gravity theories.  A general form of the Lagrangian including $A_{\mu}$ is a scalar made from $(\partial_{\nu_1} \cdots  \partial_{\nu_n} A_{\mu_1} A_{\mu_2} \cdots A_{\mu_m})$ by contracting all indices with metrics or antisymmetric tensors.  By using the dimensional analysis in Eq.(\ref{effective dimension}), we have the following dimensions for the candidates of the Lagrangian density and for the integration volume,
\begin{eqnarray}
&&[ F_{\mu\nu}F^{\mu\nu}]'=\frac{1}{[t_{\star}]^2}, ~[A_{\mu}A^{\mu}A_{\nu}A^{\nu}]'=\left(\frac{[E_{\star}]}{[t_{\star}]}\right)^2, ~[\partial_{\mu} A^{\mu}]'=\frac{1}{[t_{\star}]}, ~[R(g)]'=\frac{1}{[E_{\star}][t_{\star}]}, \nonumber \\
~&&\mathrm{and}~\int d^n x \sqrt{\det g_{\mu\nu}}=([E_{\star}][t_{\star}])^{\frac{n}{2}}.
\end{eqnarray}

Then, the action integral $I^{\star}_2$ of $\mathcal{L}^{\star}_2$ yields
\begin{eqnarray}
I^{\star}_2&=& \int d^n x \sqrt{g(x)}~ \mathcal{L}^{\star}_2=\frac{1}{(\tau k_B T)^{\frac{n}{2}}}\int d^n x \sqrt{g(x)}  \left[ -\frac{1}{4} a'_1 \tau^2 F_{\mu\nu}(x) F^{\mu\nu}(x) \right. \nonumber \\
&& +a'_2 \left(\frac{\tau}{k_B T}\right)  (A_{\mu}(x) A^{\mu}(x))
 -\frac{1}{2}  a'_3  \tau^2 \left(\partial_{\mu} A^{\mu}\right)^2 + \cdots \nonumber \\
&&\left. - b'_1 (\tau k_B T)   R(x) - b'_2 \tau^2 R_{\mu\nu}(x) A^{\mu}(x) A^{\nu}(x)  \right]+\cdots. \label{action star 2}
\end{eqnarray}

Then, the thermal fluctuation of $A_{\mu}(x)$ and $g_{\mu\nu}(x)$ can be taken into account by the following averaging,
\begin{eqnarray}
\langle O(A_{\mu}, ~g_{\mu\nu}) \rangle_{A, g} \propto \int \mathcal{D} A_{\mu}(x) \mathcal{D} g_{\mu\nu}(x) O(A_{\mu}, g_{\mu\nu}) e^{-I^{\star}_2(A_{\mu}, ~g_{\mu\nu})}.
\end{eqnarray}

Before ending this section we remark again the meaning of the two-step average of $\langle \cdots \rangle_{x}$ and $\langle \cdots \rangle_{A, g}$.  The former takes into account the effect of random forces from microscopic materials, but the macroscopic force and the transport coefficients are definitely given.  This is the first step averaging $\langle \cdots \rangle_{x}$.  If we examine more precisely, the macroscopic force and the transport coefficients given at a point $x$ in the thermal space, however, are affected by the force and the transport coefficients near $x$, that is, the mutual interactions between the forces and the transport coefficients will not be ignored.  Then, these interactions can be taken into account by the second step averaging $\langle \dots \rangle_{A, g}$.

This two-step averaging is also used in the standard treatment of a charged particle in the so-called ``proper-time formalism''\cite{proper time formalism}.  The second averaging over $A_{\mu}$ and $g_{\mu\nu}$ means the inclusion of radiative corrections to a charged particle motion by gauge bosons and gravitons.  Therefore, our formulation of two-step averaging can be useful to study the renormalization effects in thermodynamics.

\section{Entropy production by a cyclic thermodynamic process}
In this section we examine the entropy production by a ``cyclic thermodynamic process'' which depicts a closed circle $C$ in the Thermodynamic Space (ThS).

We begin with rewriting the transition probability $\Psi(\alpha, t|\alpha_0, 0)$ in Eq. (1) in our terminology.  It gives the change of the existence probability density of a state from the initial time $t=0$ to the final time $t$.  If we denote the probability density of a state at $(x^1, x^2, \cdots)$ at time  $t$ by $\phi(x, t)$, then we have
\begin{eqnarray}
\phi(x, t)=\Psi(x, t|x_0, 0)\phi(x_0, 0),
\end{eqnarray}
where $\Psi(x, t|x_0, 0)$ is given by
\begin{eqnarray}
\Psi(x, t|x_0, 0)=\int \mathcal{D} g_{\mu\nu}(x) \int \mathcal{D} A_{\mu}(x) \int_{x_0}^{x} \mathcal{D} x^{\mu}(t')~\sqrt{g(x)} ~e^{-(I^{\star}_1+I^{\star}_2)}.
\end{eqnarray}
Here the action $I^{\star}_1$ and $I^{\star}_2$ are, respectively,\footnote{It is noted that $1/2k_BT$ is factored out only for $I^{\star}_1$.}
\begin{eqnarray}
I^{\star}_1=\frac{1}{2k_BT} \int_{0}^{t} dt' ~\mathcal{L}^{\star}_1, ~\mathrm{and}~I^{\star}_2=\int d^nx  \sqrt{g(x)} ~\mathcal{L}^{\star}_2.
\end{eqnarray}

The Boltzmann's principle states that the entropy is expressed in terms of the existence probability $\phi(x, t)$:
\begin{eqnarray}
\hat{S}(x, t)= k_B \ln \phi(x, t).
\end{eqnarray}
To make clear that this entropy includes thermal fluctuations, we use the notation $\hat{S}$ with ``hat''.  This entropy with ``hat'' is given at each point in ThS, and fluctuates according to the fluctuation of the existence probability.

Then, if a cyclic thermal process $C$ is depicted as a closed circle $C$ in ThS, the entropy production after this process is over, reads
\begin{eqnarray}
\Delta_{C} \hat{S}= k_B \ln{\Psi[C]}.
\end{eqnarray}
Here $\Psi[C]$ is given for the closed path $C$ as
\begin{eqnarray}
\Psi[C]=\int \mathcal{D} g_{\mu\nu}(x) \int \mathcal{D} A_{\mu}(x)~e^{-I^{\star}_2} \times\left(e^{-\frac{1}{2k_BT} \oint_C dt \mathcal{L}^{\star}_1}\right)=\left\langle e^{-\frac{1}{2k_BT} \oint_C dt \mathcal{L}^{\star}_1} \right\rangle_{g_{\mu\nu}, ~A_{\mu}}.
\end{eqnarray}

Corresponding to fix the path $C$, $x^{\mu}(t)$ is fixed to a special one $\bar{x}^{\mu}(t)$, and the sum over different paths in ThS is abandoned. Here we also fix the kinetic coefficients $g_{\mu\nu}(x)$ and the generalized forces $A_{\mu}(x)$ at ``the most probable configurations'' (such as the classical solutions), $g_{\mu\nu}(x)=\bar{g}_{\mu\nu}(x)$ and $\bar{A}_{\mu}(x)$ in the path integrations over $g_{\mu\nu}(x)$ and $A_{\mu}(x)$, then
\begin{eqnarray}
\Delta_{C} \hat{S}&=&-\frac{1}{2T} \oint_C dt~ \mathcal{L}^{\star}_1 (\bar{x}^{\mu}, ~\bar{g}_{\mu\nu}, ~\bar{A}_{\mu}) \\
&=&-\frac{1}{2T} \oint_C dt~ \left(\frac{1}{2} \bar{g}_{\mu\nu}(\bar{x}) \dot{\bar{x}}^{\mu}\dot{\bar{x}}^{\nu} + \frac{1}{2} \bar{g}^{\mu\nu}(\bar{x}) \bar{A}_{\mu}(\bar{x}) \bar{A}_{\nu}(\bar{x})- \dot{\bar{x}}^{\mu} \bar{A}_{\mu}(\bar{x})\right). \\
&\equiv&  (\Delta_C \hat{S})_1 +  (\Delta_C \hat{S})_2 + (\Delta_C \hat{S})_3   \label{entropy production}
\end{eqnarray}

In the following discussion $A_{\mu}(x)$ and $g_{\mu\nu}(x)$ are considered to represent specific classical solutions of $\bar{A}_{\mu}(x)$ and $\bar{g}_{\mu\nu}(x)$, even if they are written without ``bar''.

There are three terms giving the ``entropy production''.\footnote{ The ``entropy'' $\hat{S}$ we are discussing is that given by the existence probability of the thermodynamic system.  It fluctuates due to the fluctuation of the thermodynamic variables.  It includes two more terms, $(\hat{S})_1$ and $(\hat{S})_2$, other than the usual classical entropy $(S)_{\mathrm{c}}=(\hat{S})_3=\frac{1}{2T} \int^t dt' \dot{x}^{\mu}(t')A_{\mu}(x(t'))$.  The classical entropy satisfies the second law of thermodynamics, $\dot{S}_{\mathrm{c}}=(\dot{\hat{S}})_3=\dot{x}^{\mu}A_{\mu}(x)= g_{\mu\nu}(x)\dot{x}^{\mu}\dot{x}^{\nu} \ge 0$, when the constitutional equation is properly chosen. }
Here, following the ``detailed fluctuation theorem''\cite{detailed fluctuation theorem} we divide the three terms of entropy production into two categories, according to the ``even and odd'' properties under the time reversal transformation $\mathcal{T}$; the time reversal exchanges a forwardly driven process to its backwardly driven one, keeping the same track $C$:
\begin{eqnarray}
\mathcal{T}: \begin{cases} t \to t'=t_a+t_b-t,~(t_a \le t \le t_b), \\
x^{\mu}(t) \to x^{\mu}(t'), \\
\dot{x}^{\mu}(t) \to -\dot{x}^{\mu}(t'), \\
\int_{t_a}^{t_b} dt \to \int_{t_b}^{t_a} dt'.
\end{cases}
\end{eqnarray}
Then, we can understand that $(\Delta_C \hat{S})_1$ and $(\Delta_C \hat{S})_2$ are $\mathcal{T}$-even (symmetric under the time reversal transformation), while $(\Delta_C \hat{S})_3$ is $\mathcal{T}$-odd (anti-symmetric under the transformation).

The irreversible process means the forward and backward processes of it differ, so that $(\Delta_C \hat{S})_3$ is responsible for the entropy production in the irreversible process.
On the other hand $(\Delta_C \hat{S})_1$ and $(\Delta_C \hat{S})_2$ are entropy production due to fluctuations in the reversible process.
Usually, the entropy production in an irreversible process is defined by the difference of $(\Delta_C \hat{S})_3$ and its time reversally transformed one.  Thus, $(\Delta_C \hat{S})_{\mathrm{irr}}=2(\Delta_C \hat{S})_3$.

Now, let us estimate $(\Delta_C \hat{S})_{\mathrm{irr}}=\frac{1}{T} \oint_C d\bar{x}^{\mu} A_{\mu}(\bar{x})$.

The third term in the right hand side of Eq.(\ref{entropy production}) has a familiar form, which counts the number of lines of magnetic flux passing through the circle $C$; by using Stokes' theorem, we have
\begin{eqnarray}
(\Delta_C \hat{S})_{\mathrm{irr}}=\frac{1}{2T} \int\!\!\!\int_{S} dx^{\mu} dx^{\nu} ~\overline{F}_{\mu\nu}(x)=\frac{1}{T} \Phi_m(C),
\end{eqnarray}
where $S$ is the surface whose boundary is $C$.
Accordingly, the entropy produced after a round trip along $C$ is given by the magnetic flux $\Phi_m(C)$ passing through the circle $C$. Without the magnetic field $F_{\mu\nu}=0$, no entropy production occurs, and the cyclic thermodynamic process along $C$ is reversible. When $\Phi_m(C)\ne 0$, the cycle of the thermodynamic process becomes irreversible.

To know the origin of this entropy production, we have to know the source of magnetic flux.  It is natural to think that the magnetic flux is generated and absorbed at magnetic monopoles.\footnote{It is no problem if monopole exists only as a combination of magnetic dipole. Furthermore, the motion of the other $x^{\mu}$s  can generate the magnetic field passing through $C$, since the coupling of $x^{\mu}(t)$ to $A_{\mu}(x)$ is the same as in electromagnetism.}  When the magnetic flux generated by these magnetic monopoles passes through the circle $C$, the irreversible entropy production occurs.

If the volume having a monopole inside is $V$, then by using Gauss' theorem, we have
\begin{eqnarray}
\int\!\!\!\int\!\!\!\int_{V} d^3 x \sum_{\mathrm{cyclic}} \partial_{\lambda} F_{\mu\nu}(x) = \frac{1}{2} \int\!\!\!\int_{S=\partial V} dx^{\mu} dx^{\nu} ~F_{\mu\nu}(x) = \Phi_m,
\end{eqnarray}
where $\sum_{\mathrm{cyclic}} \partial_{\lambda} F_{\mu\nu}(x)$ vanishes at regular points, but can be non-vanishing at a singular point $x_m$ where a monopole $m$ is located.  That is, we can assume that
\begin{eqnarray}
\sum_{\mathrm{cyclic}} \partial_{\lambda} F_{\mu\nu}(x)=\sum_{\mathrm{cyclic}} \partial_{\lambda} \left(\partial_{\mu}A_{\nu}(x)-\partial_{\nu}A_{\mu}(x) \right)=\Phi_m \delta^{(3)}(x-x_m).
\end{eqnarray}
In terms of the thermodynamics, if we choose three thermodynamic variables $(\alpha_1, \alpha_2, \alpha_3)$, the point $\alpha_S=(\alpha_1, \alpha_2, \alpha_3)$ becomes a source of entropy production, if the corresponding thermodynamic forces $(X_1(\alpha), X_2(\alpha), X_3(\alpha))$ behave singular at $\alpha_S$ where the following relation holds,
\begin{eqnarray}
\sum_{\mathrm{(1, 2, 3)cyclic}} \frac{\partial}{\partial \alpha_1} \left( \frac{\partial X_3 (\alpha)}{\partial \alpha_2 }-\frac{\partial X_2(\alpha)}{\partial \alpha_3} \right) =\Phi_m \times\delta^{(3)}(\alpha-\alpha_S).
\end{eqnarray}
The entropy production by the cyclic process $C$ is reduced by a factor $f$, even if the circle $C$ is located near the position of the monopole, since the fraction $f$ of $\Phi_m$ can pass through a thermodynamic cycle $C$, namely
\begin{eqnarray}
\Phi_m(C)=f \Phi_m, ~\mathrm{or}~(\Delta_C \hat{S})_3= 2T f \Phi_m. \label{f}
\end{eqnarray}

A rough estimation of $f$ will be given in Section 6.

Next, we will examine the other terms in Eq.(\ref{entropy production}).

The first term is $(\Delta_C \hat{S})_1=\frac{1}{4T} \oint_C dt~ \bar{g}_{\mu\nu}(\bar{x})\dot{\bar{x}}^{\mu}\dot{\bar{x}}^{\nu}$.  This term is related to the curvature of the ThS, and does not exist in the classical theory.  The estimation of this is also a standard one.  Let $x_0$ be the position of the center of mass of the circle $C$; $\oint_C dt (\bar{x}^{\mu}(t)-x^{\mu}_0)=0$ for all $\mu$.  The $x_0$ is usually not on the curve $C$. Any point $\bar{x}$ on $C$ can be connected to $x_0$ by a geodesic curve. If the distance between $x_0$ and $\bar{x}$ is chosen to be the geodesic distance $s$ between them; $s=\int_{x_0}^{\bar{x}} ds \sqrt{g_{\mu\nu}\dot{x}^{\mu}\dot{x}^{\nu}}$, then we have the coordinate system called ``geodesic normal coordinates''.  Using this coordinate system, the metric can be expressed as \cite{Brevin}:
\begin{eqnarray}
g_{\mu\nu}(x)=g_{\mu\nu}(x_0) - \frac{1}{3} R_{\mu\alpha\nu\beta}(x_0)(x-x_0)^{\alpha}(x-x_0)^{\beta} + \cdots,
\end{eqnarray}
where $R_{\mu\alpha\nu\beta}$ is the Riemann tensor which measures how the space around the circle $C$ is curved.  If the curvature is small and the above expansion is allowed, then we can express the entropy production by the first order term of the Riemann tensor:
\begin{eqnarray}
(\Delta_C \hat{S})_1&=&\frac{1}{4T} \left\{ g_{\mu\nu}(x_0) \oint_C dt~ \dot{\bar{x}}^{\mu}\dot{\bar{x}}^{\nu} -\frac{1}{3} R_{\mu\alpha\nu\beta}(x_0)\oint_C dt~\dot{\bar{x}}^{\mu} (\bar{x}-x_0)^{\alpha} \dot{\bar{x}}^{\nu} (\bar{x}-x_0)^{\beta}\right\} \nonumber \\
&=& \frac{1}{4T}  \left\{ g_{\mu\nu}(x_0) a^{\mu\nu}(C) - R_{\mu\alpha\nu\beta}(x_0) a^{\mu\alpha\nu\beta} (C)\right\},
\end{eqnarray}
where
\begin{eqnarray}
&&a^{\mu\nu}(C)=\oint_C dt~ \dot{\bar{x}}^{\mu}\dot{\bar{x}}^{\nu}, \\
&&a^{\mu\alpha\nu\beta}(C) =\oint_C dt~\dot{\bar{x}}^{\mu} (\bar{x}-x_0)^{\alpha} \dot{\bar{x}}^{\nu} (\bar{x}-x_0)^{\beta}.
\end{eqnarray}

Thus, if ThS is curved around the curve $C$ giving the thermodynamic process, the entropy production is generated even in the reversible process by the curvature of the space $R_{\mu\alpha\nu\beta}$. This means the thermodynamic process can measure the curvature of ThS.  In  later sections  we will consider what is the source of this entropy production.

The remaining is the entropy production by the second term, which is
\begin{eqnarray}
(\Delta_C \hat{S})_2=\frac{1}{4T} \oint_C dt~ \bar{g}^{\mu\nu}(\bar{x})\bar{A}_{\mu}(\bar{x})\bar{A}_{\nu}(\bar{x})=\frac{1}{2T}\oint_C dt~ \Phi^{(-1)}(A(\bar{x}(t))).
\end{eqnarray}

To understand the difference between this term $(\Delta_C \hat{S})_2$ and the other terms, we will use the property that three types of the entropy production have different dependency on $\dot{x}^{\mu}$.
Let us introduce the time (period) $P$ being used to go around the same circle $C$.  We will scale time uniformly to describe the slow and the rapid operation of the thermodynamic processes.

Then, we know
\begin{eqnarray}
(\Delta_C \hat{S})_1 \propto \frac{1}{P}, ~~(\Delta_C \hat{S})_2 \propto P, ~\mathrm{and}~ (\Delta_C \hat{S})_3 \propto 1.
\end{eqnarray}

 From this we understand that $(\Delta_C \hat{S})_1$ dominates for the rapid operation of going around the circle $C$ and $(\Delta_C \hat{S})_2$ is dominant for the slow operation, or it remains even for the quasi-equilibrium process.  These are time reversal even entropy productions.  The third term $(\Delta_C \hat{S})_3$ is time reversal odd, and furthermore it does not depend on the period $P$, that means this entropy production is not given dynamically, but topologically in terms of topological quantum numbers such as the monopole charge.

The above discussion shows that we can separate three types of entropy production experimentally by controlling the rapid and slow operations of thermodynamic processes.

\section{ The effective action induced by the non-equilibrium thermodynamic action  }

In the last section we consider the ``cyclic thermodynamic process'' which depicts a closed circle $C$ in the Thermodynamic Space (ThS).   We consider this ``thermodynamic cycle'' to be a tiny probe of (or a small perturbation to) the thermodynamic space, so that it does not contribute to $\mathcal{L}^{\star}_2$, or the whole dynamics of $g_{\mu\nu}$ and $A_{\mu}$.  However, the virtually arising infinite number of ``circles" do contribute to the dynamics of $g_{\mu\nu}$ and $A_{\mu}$. The infinite number of these virtual processes are known to generate the so-called ``effective action'', which contributes to $\mathcal{L}^{\star}_2$.  Therefore, in this section we derive this effective action $I^{\star}_{\mathrm{eff}}=\int dx \sqrt{g(x)} \mathcal{L}^{\star}_{\mathrm{eff}}$.

We start from the Onsager-Machlup and Hashitsume (OMH) Lagrangian, which is in our description
\begin{eqnarray}
\mathcal{L}^{\star}_1=\frac{1}{2} g_{\mu\nu}(x) \left(\dot{x}^{\mu}(t)-A^{\mu}(x) \right) \left(\dot{x}^{\nu}(t)-A^{\nu}(x) \right).
\end{eqnarray}

The virtual fluctuation of $x^{\mu}(t)$ occurs everywhere in ThS, the sum of all these virtual fluctuations gives an effective action for $g_{\mu\nu}(x)$ and $A_{\mu}(x)$.  What we are going to estimate is this effective action $I^{\star}_{\mathrm{eff}, \; 1}$, or the effective Lagrangian $\mathcal{L}^{\star}_{\mathrm{eff}, \; 1}(x) $, being induced by $\mathcal{L}^{\star}_1(t) $:
\begin{eqnarray}
&&e^{-I^{\star}_{\mathrm{eff}, 1}}=e^{-\frac{1}{2k_B T} \int d^{N}x ~\mathcal{L}^{\star}_{\mathrm{eff}, 1}(x)} \\
&& \equiv \int \mathcal{D} x^{\mu}(t) e^{-\frac{1}{2k_B T} \int dt \mathcal{L}^{\star}_1(t) }.
\end{eqnarray}
The virtual fluctuations (or excitations) occur disconnectedly (as disconnected circles $C$s) depicted by the trajectory of $x^{\mu}(t)$ at different place, but the sum over all these virtual fluctuations can be summed up to an exponential with a power coming from a single connected fluctuation (a single circle $C$) at a single place.  We denote this single circle by $C(x_0, s)$, where $x_0$ denotes the center of mass of the trajectory, and $s$ is a period of the circular motion, $0 \le t \le s$. That is, we have
\begin{eqnarray}
I^{\star}_{\mathrm{eff}, \; 1}=-\int_0^{\infty} \frac{ds}{s} \int_{C(x_0, s)} \mathcal{D} x^{\mu}(t) e^{-\frac{1}{2k_B T} \int_0^s dt \mathcal{L}^{\star}_1(t) },
\end{eqnarray}
where $\frac{1}{s}$ in the integral takes off the degeneracy relating to the ambiguity of the starting point $x(0)$, when summing over the period of the circle.

The derivation of the effective action is similar to that of Heisenberg-Euler formula in QED, and hence we follow the derivation of H-E formula \cite{proper time formalism}.
Corresponding to the circle $C(x_0, s)$, we expand $g_{\mu\nu}(x)$ and $A_{\mu}(x)$ around $x_0$,
\begin{eqnarray}
\begin{cases}
~g_{\mu\nu}(x)=g_{\mu\nu}(x_0)-\frac{1}{3} R_{\mu\alpha\nu\beta}(x_0) (x-x_0)^{\alpha}  (x-x_0)^{\beta}+ \cdots, \\
~A_{\mu}(x)=A_{\mu}(x_0)+(x-x_0)^{\nu} \partial_{\nu} A_{\mu} (x_0)+ \cdots,
\end{cases}
\end{eqnarray}
and the closed circle $C(s, x_0)$ is represented by\footnote{Here we use the notation $\dagger$, but this is nothing but the complex conjugation of the complex Fourier expansion parameters $a^{\mu}_n$.}
\begin{eqnarray}
x^{\mu}=x^{\mu}_0 + \frac{1}{\sqrt{s}} \sum_{n=1}^{\infty} \left( a^{\mu}_n e^{-2\pi i nt/s}+ a^{\mu \dagger}_n e^{2\pi i nt/s} \right).
\end{eqnarray}

Then, we have
\begin{eqnarray}
 \int_{0}^{s} dt~\mathcal{L}^{\star}_1(t) =\frac{1}{2} A_{\mu}(x_0)A^{\mu}(x_0)
+ \sum_{n=1}^{\infty} a^{\mu \dagger}_n \ell_{\mu\nu}(n) a^{\nu}_n.
\end{eqnarray}
where
\begin{eqnarray}
\ell_{\mu\nu}(n) &=& (2 \pi n/s)^2 g_{\mu\nu}(x_0) + i(2\pi n/s) \left(\partial_{\mu} A_{\nu}(x_0)- \partial_{\nu} A_{\mu}(x_0) \right) \nonumber \\
&+&2 \left( \partial_{\mu}A_{\rho}(x_0) \partial_{\nu}A^{\rho}(x_0) \
- \frac{1}{3} R_{\alpha \mu \beta \nu}(x_0) A^{\alpha}(x_0) A^{\beta}(x_0) \right),
\end{eqnarray}
and the raising and lowering of indices are done by $g_{\mu\nu}(x_0)$.

Under this mode expansion, the path integral over $x^{\mu}(t)$ becomes the mode integrations:
\begin{eqnarray}
S^{\star}_{\mathrm{eff}, \; 1}=-\int d^N x_0 \sqrt{g(x_0)} \int_0^{\infty} \frac{ds}{s} e^{-\frac{1}{2k_BT} m^2 s}  \prod_{n=1}^{\infty} \int da^{\mu}_n \; da^{\mu\dagger}_n ~ e^{-\frac{1}{2k_BT} a^{\mu \dagger}_n \ell_{\mu\nu}(n) a^{\nu}_n},
\end{eqnarray}
where $m^2 \equiv \frac{1}{2} A_{\mu}(x_0)A^{\mu}(x_0)$ plays the role of a mass of ``particle'' moving along the trajectory $x^{\mu}(t)$ with ``proper time'' $t$.

Here, we introduce the matrices $\bm{F}$ and $\bm{G}$ for the vector fields (affinity forces), the metric (kinetic constants) and the curvature, which are ``effectively induced'' by the thermodynamic action of OMH:
\begin{eqnarray}
\begin{cases}
~\bm{F}_{\mu}^{~\nu}=\partial_{\mu} A^{\nu}(x_0)- \partial^{\nu} A_{\mu}(x_0), \\
~\bm{G}_{\mu}^{~\nu}=\partial_{\mu}A_{\rho}(x_0) \partial^{\nu}A^{\rho}(x_0) \
- \frac{1}{3} R_{\alpha \mu \beta}^{~~~~\nu}(x_0) A^{\alpha}(x_0) A^{\beta}(x_0),
\end{cases}
\end{eqnarray}
the matrix $\bm{\ell}(n)$, $\bm{\ell}_{\mu}^{~\nu}(n)=\ell_{\mu}^{~\nu}(n)$ has the following expression:
\begin{eqnarray}
\tilde{\bm{\ell}} (n) \equiv \frac{\bm{\ell}(n)\vert_{F, G \ne 0}}{\bm{\ell}(n)\vert_{F, G = 0}}= \frac{\bm{\ell}(n)}{(2 \pi n/ s)^2}= \bm{1} + \frac{i}{2} \left(s/\pi n\right) \bm{F} + \frac{1}{2} \left(s/\pi n\right)^2 \bm{G}.
\end{eqnarray}

Then, summing over all the possible virtual fluctuations, we have
\begin{eqnarray}
\mathcal{L}^{\star}_{\mathrm{eff},\; 1}= -\int_0^{\infty} \frac{ds}{s}~ \Psi(x', s |x', 0)\vert_{F, G=0} \times e^{-m^2 s} \prod_{n=1}^{\infty} \frac{1}{\mathrm{det} ~\tilde{\bm{\ell}}(n)},
\end{eqnarray}
where $\Psi(x', s|x', 0)\vert_{F, G=0}$ is the transition amplitude in the free case without induced fields $\bm{F}$ and $\bm{G}$, and the expression can be consistent with the free case by using $\tilde{\bm{\ell}} (n)$.  (See \cite{proper time formalism}.)  The $\Psi$ can be found in Appendix B on the Fokker-Planck equation as $\Psi(x', s |x', 0)\vert_{F, G=0}=1/(4\pi k_B T~s)^{N/2}$, where the dimension of ThS is denoted here by $N$.  Using $\mathrm{det}(\tilde{\bm{\ell}})= e^{\mathrm{Tr} \ln \tilde{\bm{\ell}}}$, we obtain the effective Lagrangian induced by the thermodynamic action as follows:
\begin{eqnarray}
\mathcal{L}^{\star}_{\mathrm{eff},\; 1}(x)= \frac{-1}{(4\pi k_B T)^{N/2}}\int_0^{\infty} \frac{ds}{s^{\frac{N}{2}+1}} e^{-\frac{1}{2k_BT} m^2 s} e^{-\sum_{n=1}^{\infty}\mathrm{Tr} \ln \tilde{\bm{\ell}}(n)},
\end{eqnarray}
where
\begin{eqnarray}
&& \mathrm{Tr} \ln \tilde{\bm{\ell}}(n)=\mathrm{Tr} \ln \left( \bm{1} + \frac{i}{2} (s/\pi n) \bm{F} (x) + \frac{1}{2} (s/\pi n)^2 \bm{G}(x) \right) \\
&&= \mathrm{Tr} \left[ \frac{1}{2}\left\{ i (s/\pi n) \bm{F}(x) + (s/\pi n)^2 \bm{G}(x) \right\} - \frac{1}{8} \left\{ i (s/\pi n) \bm{F}(x) + (s/\pi n)^2 \bm{G}(x) \right\}^2 + \cdots \right],~~~~~~~
\end{eqnarray}
where $x$ stands for the original $x_0$.

The first few terms of the effective Lagrangian in the ThS induced by the thermodynamic action of Onsager-Machlup and of Hashitsume are
\begin{eqnarray}
&&\mathcal{L}^{\star}_{\mathrm{eff},\; 1}(x) = \frac{-1}{(4\pi k_B T)^{\frac{N}{2}}} \nonumber \\
&& \times \left[\frac{\zeta(2)}{\pi^2} \Gamma\left(2-\frac{N}{2}\right) \left(\frac{m^2}{2k_BT}\right)^{\frac{N}{2}-2}\left\{ \frac{1}{2} (\partial_{\mu}A_{\nu}(x))^2-\frac{1}{3} R_{\mu\nu} (x) A^{\mu}(x)A^{\nu}(x) -\frac{1}{8} F_{\mu\nu}F^{\mu\nu}(x) \right\} \right. \nonumber \\
&& \left. - \frac{\zeta(4)}{8\pi^4} \Gamma \left(4-\frac{N}{2}\right) \left(\frac{m^2}{2k_BT}\right)^{\frac{N}{2}-4}  \biggl\{  \partial_{\mu}A_{\lambda}(x) \partial^{\nu} A^{\lambda}(x) \partial_{\nu}A_{\rho}(x) \partial^{\mu} A^{\rho}(x)    \right. \nonumber \\
&&\left. \left. -\frac{2}{3} R^{\alpha\mu\beta\nu}(x) A_{\alpha}(x) A_{\beta}(x)\partial_{\mu}A_{\rho}(x) \partial_{\nu} A^{\rho}(x) +\frac{1}{9} {R_{\alpha\mu\beta}}^{\nu}(x) {R_{\alpha'\nu\beta'}}^{\mu}(x) A^{\alpha}(x) A^{\beta}(x) A^{\alpha'}(x) A^{\beta'}(x)  \right\} \right],~~~~~~
\end{eqnarray}
where $m^2=\frac{1}{2}A_{\mu}(x)A^{\mu}(x)$, $\zeta(2)=\frac{\pi^2}{6}$,  $\zeta(4)=\frac{\pi^4}{90}$, and $\Gamma(z)=\int_0^{\infty} ds~ s^{z-1}e^{-s}$.

Therefore, we understand that the action like $\mathcal{L}^{\star}_2$ can be induced by the thermodynamic action of OMH through the virtual fluctuations or excitations.  This gives an motivation and an candidate for the additional action $\mathcal{L}^{\star}_2$ which gives a weight to find the most probable configuration for $g_{\mu\nu}(x)$ and $A_{\mu}(x)$.

To give an example in which our gravity analog model works, we will study $N=4$ case.   In this case the effective action diverges in the ultraviolet (UV) region, giving a pole at $N=4$.  If we choose $N$ a little smaller than 4 with a deviation  $\varepsilon=2-\frac{N}{2} > 0$:
\begin{eqnarray}
\Gamma\left(2- \frac{N}{2} \right) \left(\frac{m^2}{2 k_B T} \right)^{\frac{N}{2}-2} \xrightarrow[N \to 4]{} \frac{1}{\left(2-\frac{N}{2}\right)} + \mathrm{finite~contributions} \approx \frac{1}{\varepsilon}.
\end{eqnarray}
Then, the effective Lagrangian becomes
\begin{eqnarray}
\mathcal{L}^{\star}_{\mathrm{eff}, \;1}= \frac{-1}{6 \varepsilon (4\pi k_B T)^{2}}  \left\{ \frac{1}{2} (\partial_{\mu}A_{\nu}(x))^2-\frac{1}{3} R_{\mu\nu} (x) A^{\mu}(x)A^{\nu}(x) -\frac{1}{8} F_{\mu\nu}F^{\mu\nu}(x) \right\}. \label{a candidate for L2}
\end{eqnarray}

If we compare $\mathcal{L}^{\star}_{\mathrm{eff}}$ to $\mathcal{L}^{\star}_2$ in Eq.(\ref{action star 2}), we know that the derived effective action consists of three terms with coefficients $a_1', a_3'$, and $b_2'$ in Eq.(\ref{action star 2}), and the overall coefficient reproduces that derived by the dimensional analysis, $\tau^2/(\tau k_B T)^{N/2} \to 1/(k_B T)^2$ for $N \to 4$.  Therefore, the effective action $\mathcal{L}^{\star}_{\mathrm{eff}, \;1}$ is a good candidate for $\mathcal{L}^{\star}_2$.\footnote{If the non-equilibrium thermodynamics is a topological theory, then virtual contribution from the thermodynamic variables $x^{\mu}(t)$ is cancelled by that from the anti-commuting ghost and anti-ghost fields, $c^{\mu}(t)$ and $\bar{c}_{\mu}(t)$. The physical implication of the ghost and anti-ghost fields is, however, not clear so far.  Therefore, we take a stance that the treatment by Onsager-Machlup and Hashitsume without ghost fields is physically acceptable, and hence we ignore the contribution of the ghosts in the above estimation of the effective action.}

\section{Magnetic flux and curvature in Thermodynamic Space (ThS)}

The source of the magnetic flux is attributed to monopoles in  Section 5, and they contribute to the entropy production in the irreversible processes of the time reversal odd. Here we examine how the magnetic flux expands and decays.  To see this effect we have to consider the kinetic energy of the field $A_{\mu}(x)$ in $I^{\star}_2$.  There are a number of candidates, but as was understood in the above, the entropy production is given by the magnetic field $F_{\mu\nu}$ which comes from the transverse component of the field $A^{\mu}$, not from the longitudinal component $\partial^{\mu}A_{\mu}$.  Therefore, if the space is not curved, the candidate action up to the second order in $A$ is
\begin{eqnarray}
(I^{\star}_2)_A&=&\int d^n x \sqrt{g(x)}~ \mathcal{L}^{\star}_2=\frac{1}{(\tau k_B T)^{\frac{n}{2}}}\int d^n x \sqrt{g(x)}  \left[ -\frac{1}{4} a'_1 \tau^2 F_{\mu\nu}(x) F^{\mu\nu}(x) \right. \nonumber \\
&& \left. +a'_2 \left(\frac{\tau}{k_B T}\right) A_{\mu}(x) A^{\mu}(x) \right].
\end{eqnarray}

The equation of motion (most probable configuration) reads
\begin{eqnarray}
\nabla^{\nu} F_{\nu\mu}(x)- \frac{2a'_2}{a'_1} \left(\frac{1}{k_BT}\right) A_{\mu}=0,
\end{eqnarray}
where $\nabla^{\nu}$ is the covariant derivative in the curved space.  In the above equation of motion we have not included the source of the current $j^{\mu}(x)$ for $A_{\mu}$, since we consider that the thermodynamic cycle, or the motion of $x^{\mu}(t)$, is a tiny probe of ThS and does not contribute to $j^{\mu}(x)$.

What is necessary here is not this equation of motion\footnote{More rigorously, the additional magnetic field around monopole appears which, however, damps by the mass\cite{Ignatiev}.}, but is the Bianchi identity and its violation by monopole; the monopole is assumed to  locate at $x_m$ with the magnetic charge $\Phi_m$.   The source of monopole is difficult to introduce in principle, so that it is  introduced as a violation of the Bianchi identity, or the singularity of the vector field $A_{\mu}(x)$ here.\footnote{This is the Dirac's method in the Abelian gauge theory. In the non-Abelian gauge theory, the monopole can be introduced without singularities as a classical solution of the action.  There is another method by D. Zwanziger, that is able to introduce the monopole current in the action.}  Then we have
\begin{eqnarray}
\sum_{\mathrm{cyclic}} \nabla_{\lambda} F_{\mu\nu}(x)=\sum_{\mathrm{cyclic}} \nabla_{\lambda} \left(\nabla_{\mu}A_{\nu}(x)-\nabla_{\nu}A_{\mu}(x) \right)=-\Phi_m \delta^{(n)}(x-x_m),
\end{eqnarray}

Here, we consider the case of $n=3$ (three-dimensional thermodynamic space).  Then, the magnetic field $\bm{B}=-(F_{23},~F_{31},~F_{12})$ behaves as usual
\begin{eqnarray}
\bm{B}(\bm{x}) =\overline{\bm{B}}(\bm{x}) =\frac{\Phi_m}{4\pi} \frac{\bm{x}-\bm{x}_m}{\vert \bm{x}-\bm{x}_m \vert^3}.
\end{eqnarray}

Therefore, given the solid angle $\Omega$ of looking $C$ from the position of the monopole $x_m$, the fraction given in Eq.(\ref{f}) reads
\begin{eqnarray}
f=\Omega/4\pi.
\end{eqnarray}

Next, we examine the source of the curvature in ThS.

As is well-known in general relativity, the presence of (very) massive bodies or their (rapid) motion deforms the flat space to a curved space; the degree of this deformation of space (or the curvature of the space) can be detected by the bending of light path, or by the gravitational lens effect.  In our thermodynamic model, the kinetic coefficients give the metric $g_{\mu\nu}(x)$ of ThS.  If the metric is space independent, ThS is flat.  If the metric depends on the position, however, the curvature may appear.  In the last section this curvature can produce
the entropy in the rapidly moving reversible processes. To understand the source of this entropy production, we examine the equation of motion (Einstein equation) for the metric, which gives,
\begin{eqnarray}
R_{\mu\nu}(x) -\frac{1}{2} g_{\mu\nu}(x) R(x) = \kappa T_{\mu\nu}(x),
\end{eqnarray}

where
\begin{eqnarray}
\kappa=(b'_1)^{-1} (\tau k_B T)^{\frac{n}{2}-1},
\end{eqnarray}

and the Ricci tensor and the scalar curvature are defined by the Riemann curvature as
\begin{eqnarray}
R_{\mu\nu}=g^{\alpha\beta}R_{\alpha\mu\beta\nu}, ~\mathrm{and}~ R=g^{\mu\nu}R_{\mu\nu}.
\end{eqnarray}
 This is the equation of motion for $\mathcal{L}^{\star}_2$ in Eq.(\ref{action star 2}) about $g^{\mu\nu}(x)$.  The action in the non-equilibrium thermodynamics is not necessarily Einstein's one, and hence we combine all the contributions other than Einstein's one to $\mathcal{L}^{\star}_{\mathrm{others}}$.
Then, the energy momentum tensor $T_{\mu\nu}$ comes from $\mathcal{L}^{\star}_{\mathrm{others}}$, and is given by
\begin{eqnarray}
T_{\mu\nu}(x)= \frac{2}{\sqrt{g(x)}} \frac{\delta (\sqrt{g(x)} \mathcal{L}_{\mathrm{others}})}{\delta g^{\mu\nu}(x)}.
\end{eqnarray}
As a trial, let us ignore the other contributions $\mathcal{L}^{\star}_{\mathrm{others}}=0$, then we have the ordinary equation of motion for $g^{\mu\nu}$,
\begin{eqnarray}
R_{\mu\nu}=0.
\end{eqnarray}


The metric of thermodynamics is, however, Euclidean-like, that is its eigen-values are all positive definite, since the kinetic constants $L_{ij}$ should be a positive definite matrix in order to describe the diffusion.  Therefore,  the candidates responsible for the entropy production are ``gravitational instantons''.  A lot of examples are known.\cite{gravitational instanton}

A simple example is the Euclidean version of Schwarzschild black hole which is appropriate to analyze. We choose $n=4$ and restrict to the four-dimensional subspace $(w, x, y, z)$ in the thermodynamic space, which does not include time $t$.  The four coordinates are assumed to have the same dimensions.  If not, we should modify the variables.  Then, the metric can be given by
\begin{eqnarray}
ds^2=\left(1-\frac{2M}{r} \right)dw^2+\left(1-\frac{2M}{r} \right)^{-1} dr^2+r^2(d\theta^2 + \sin \theta^2 d \varphi^2),
\end{eqnarray}
where the polar coordinates $(r, \theta, \varphi)$ are chosen for the three-dimensional space $(x, y, z)$ and $M$ is the ``mass'' (or ``Schwarzschild radius'') of the black hole. In this Euclidean black hole, the sign in front of $dw^2$ differs from the usual Minkowsky black hole.  The solution has a curvature singularity at $r=0$, which is expected to occur through the interaction between the metric (the kinetic constants) $g_{\mu\nu}(x)$ and the thermodynamic forces $A_{\mu}(x)$.  The effect of interactions is summarized into a single parameter $M$ in the spherically symmetric solution. The explanation of $M$ in terms of the non-equilibrium thermodynamics terminology is necessary, but is, however, beyond the present scope of ours.

The entropy production $\Delta_ C \hat{S}_1$ consists of two terms. Here, we fix the curve $C$ as $(\bar{w}, \bar{r}, \bar{\theta})$ are constant, and only $\varphi$ changes as $\bar{\varphi}=\varphi(t)$ and depicts a closed curve $C$.
The first contribution reads
\begin{eqnarray}
\oint dt ~g_{\mu\nu}(x)\dot{\bar{x}}^{\mu}\dot{\bar{x}}^{\nu}=\oint dt~g_{\varphi\varphi}\dot{\varphi}^2=\bar{r}^2 \sin^2 \bar{\theta} \oint dt~\dot{\varphi}^2.
\end{eqnarray}
The solid angle $\Omega$ of looking the circle from the origin is $4\pi\sin^2{\bar{\theta}}$, so that $\sin^2 \bar{\theta}= \Omega/4\pi$.

  It is interesting to estimate the second term.  It reads
\begin{eqnarray}
&&\oint dt ~R_{\mu\nu\lambda\rho}\dot{\bar{x}}^{\mu}(t)\bar{x}^{\nu}(t)\dot{\bar{x}}^{\lambda}(t) \bar{x}^{\rho}(t) \nonumber \\
&&=\oint dt ~R_{\varphi r \varphi r} \times (r\sin \theta \dot{\varphi})^2 \times \left(\left(1-\frac{2M}{r} \right)^{-\frac{1}{2}} r\right)^2,
\end{eqnarray}
where a component of the Riemann tensor can be read from \cite{gravitational instanton}:
\begin{eqnarray}
R_{\varphi r \varphi r}=-\frac{M}{r^3}.
\end{eqnarray}
Thus we obtain
\begin{eqnarray}
\oint dt ~R_{\mu\nu\lambda\rho}\dot{\bar{x}}^{\mu}(t)\bar{x}^{\nu}(t)\dot{\bar{x}}^{\lambda}(t) \bar{x}^{\rho}(t) =-\frac{M \bar{r}^2 \sin^2 \bar{\theta}}{\bar{r}-2M}\oint dt~\dot{\varphi}^2
\end{eqnarray}
The first contribution is proportional to $\bar{r}^2$, while the second one is proportional to $\bar{r}$ if the path of the thermodynamic trajectory $C$ is far from the position of the horizon at $\bar{r}=2M$ and is less dominant. If the trajectory $C$ is, however, located in the neighborhood of the horizon, the second term becomes extremely large.

The multi-Taub-NUT solution of the Euclidean Einstein equation is known\cite{gravitational instanton}, giving
\begin{eqnarray}
ds^2=V (dw+\bm{\omega} \cdot \bm{dx} )^2 + V^{-1} \bm{dx}\cdot \bm{dx},
\end{eqnarray}
where
\begin{eqnarray}
&&V^{-1}= 1+ \frac{2\pi n_i}{\vert \bm{x}- \bm{x}_i \vert},  \\
&&\mathrm{and}~~ \bm{\nabla} \times \bm{\omega}=\bm{\nabla} V,
\end{eqnarray}
where the so-called ``nut'' is located at $\bm{x}_i$ with a strength $n_i$.  In gauge theory the ``nut'' corresponds to a self-dual ``dyon'', for which the magnetic field radiated is equal to the electric field radiated, $\bm{B}=\bm{\nabla} \times \bm{\omega}=\bm{\nabla} V=\bm{E}$.  Therefore, the discussion of the entropy production by the multi-Taub-NUT solution may resemble that of monopole in the last section for the other entropy production, $\Delta_C\hat{S}_3$.

\section{ An example:~The gravity analog model applied to a chemical reaction in a solvent }
To obtain the better understanding of our gravity analog model, a proper example of it  is desired.  We will give such an example (the OUJ model) in a chemical reaction, in which two chemical substances interact in a solvent of the van der Waals fluid/gas.  The van der Waals solvent gives a black-hole like behavior to the metric (the kinetic constant), while the oscillatory behavior between two chemical substances gives a monopole like behavior to the vector fields (the thermal forces).

We denote the numbers of molecules of two chemical substances as $Y$ and $Z$, and put them into a solvent, having $N$ molecules $(Y, Z \ll N)$ and temperature $T$ and volume $V$.  $X$ is assumed to be constant in time.  In this setting, the thermal variables are $\{x^{\mu}\}=\{T, V, Y,  Z \}$, and the thermodynamic space (ThS) can be a familiar four-dimensional one, $\{x^{\mu}\}~(\mu=0, 1, 2, 3)$.

The thermodynamic forces, corresponding to the thermal variables, can be extracted as coefficients in the Gibbs relation for the Helmholtz free energy $F$:
\begin{eqnarray}
d F = -S \; dT - p \; dV + \mu_{Y} \; dY +  \mu_{Z} \; dZ.
\end{eqnarray}
That is, the thermodynamic forces $A_{\mu}(x)$ are
\begin{eqnarray}
A_0=-S, ~A_1=-p, ~A_2=\mu_Y, ~A_3=\mu_Z.
\end{eqnarray}

In our gravity analog model, the metric (kinetic constants) $g_{\mu\nu}(x)$ and the vector field (thermodynamic forces) $A_{\mu}(x)$ are controlled by the field theoretical action $I^{\star}_2$ with Lagrangian $\mathcal{L}^{\star}_2$.  We choose $\mathcal{L}^{\star}_2=\mathcal{L}^{\star}_{\mathrm{eff}, \; 1}$ which is derived effectively in Eq.(\ref{a candidate for L2}) from the non-equilibrium thermodynamic action of OMH.  That is
\begin{eqnarray}
\mathcal{L}^{\star}_{2} &=& \frac{-1}{6\varepsilon (4\pi k_B T)^{2}}  \left\{ \frac{1}{2} (\partial_{\mu}A_{\nu}(x))^2-\frac{1}{3} R_{\mu\nu} (x) A^{\mu}(x)A^{\nu}(x) -\frac{1}{8} F_{\mu\nu}F^{\mu\nu}(x) \right\}, \label{L2 for N=4} \\
&=&\frac{-1}{6 \varepsilon (4\pi k_B T)^{2}}  \left\{-\frac{1}{3} R_{\mu\nu} (x) A^{\mu}(x)A^{\nu}(x) + \frac{1}{8} G_{\mu\nu}G^{\mu\nu}(x) \right\}, \label{L2 for N=4}
\end{eqnarray}
where $G_{\mu\nu}(x)=\nabla_{\mu} A_{\nu}(x) + \nabla_{\nu} A_{\mu}(x)$ with a covariant derivative $\nabla_{\mu}$ defined in a curved space, while $F_{\mu\nu}(x)=\nabla_{\mu} A_{\nu}(x) - \nabla_{\nu} A_{\mu}(x)= \partial_{\mu}A_{\nu}(x) - \partial_{\nu} A_{\mu}(x)$.

We will assume the first two components of the thermodynamic forces as those of the van der Waals model, and the last two components as those of the chemical potentials necessary to realize the oscillatory motion in the reaction and diffusion system:
\begin{eqnarray}
A_0(T, V) = -S(T, V), \; A_1(T, V)=-p(T, V), \; A_2(Y, Z)=\mu_Y(Y, Z), \;  A_3(Y, Z)=\mu_Z(Y, Z),
\end{eqnarray}
where $S$ and $p$ are the entropy and pressure of the solvent liquid, respectively, and $\mu_Y$ and $\mu_Z$ are chemical potentials of two chemical substances.

The entropy and the pressure are chosen to those of the van der Waals  model, and the chemical potentials are chosen as follows:
\begin{eqnarray}
\begin{cases}
-S= -Nk_B \ln (V/N-b) - N k_B \; c_v \left(1+\ln(T/T_0) \right), ~
-p=-\frac{k_B T}{V/N- b}+ a (V/N)^{-2},  \\
\mu_Y=- \frac{g_m}{2N}\; Z, ~\mu_Z= \frac{g_m}{2N} \; Y,  \label{choice of the vector fields}
\end{cases}
\end{eqnarray}
where $g_m$ can be a ``monopole charge'' which determines the magnitude of the magnetic field $F_{23}=\partial_2 A_3- \partial_3 A_2= g_m/N$ in the $(Y, Z)$ space.

The above equation for $p$ gives the equation of state of the van der Waals liquid/gas:
\begin{eqnarray}
\left\{p+a (N/V)^2\right\} \cdot \left( V/N-b\right)=k_B T,
\end{eqnarray}
where $a (>0)$ and $b (>0)$ are parameters expressing the attractive force between molecules and the finite size of them, and $c_v$ is a numerical constant to count the degrees of freedom ($c_v=3/2$ for the mono-atomic molecule).
We will determine the metric $g_{\mu\nu}(x)$ as a classical solution of the effective action $I^{\star}_2$ for the given vector fields $A_{\mu}(x)$ as in Eq.(\ref{choice of the vector fields}).

Our ansatz of the metric is
\begin{eqnarray}
ds^2 = f(T, V)^2 \; (dT^2 + dV^2) + f'(Y, Z)^2 \; (dY^2+ dZ^2) ,
\end{eqnarray}
which means the four-dimensional thermodynamic space (ThS), $M_4(T, V, Y, Z)$, is the product of two 2-dimensional spaces, $M_4=M_2(T, V) \times M_2'(Y, Z)$. (See Appendix A in which the preliminaries for this section is given.)

\underline{1) In the close-packing limit of the fluid} \\
Given the thermodynamic forces in $M_2$ are those of the van der Waals liquid/gas, then the metric of $M_2$ becomes in the close-packing limit as
\begin{eqnarray}
 g^{00}(V)=g^{11}(V) \approx ~~ C' (V-Nb)^{6},
 \end{eqnarray}
where $C'$ is a numerical constant.  The metric becomes singular at the close-packing limit of $V \to Nb$.  This metric singularity resembles  that of black-hole.  More detailed analysis is necessary to elucidate the correlation between the singularities in thermodynamics or its critical behavior and  the metric singularities in our gravity analog model. This topic will be studied in the forthcoming paper \cite{forthcoming paper}.

 \underline{2) Chemical oscillation between two substances} \\
On the other hand, if we take a monopole like configuration for the relevant thermodynamic forces in $M'_2$,
\begin{eqnarray}
A_Y(Y, Z)=\mu_Y(Y, Z)=- \frac{g_m}{2N}\; Z, ~A_Z(Y, Z)=\mu_Z(Y, Z)= \frac{g_m}{2N} \; Y.
\end{eqnarray}
The non-zero magnetic field appears on $(Y, Z)$, $F_{23}=g_m/N \ne 0$, and the metric of the space $M'_2$ becomes
\begin{eqnarray}
g^{22}(Y, Z)=g^{33}(Y, Z)=(1/F(r))^2= K \left\{ Y^2+Z^2 \right\}^{\sqrt{3}-1}.
\end{eqnarray}

Then, the temporal change of the chemical substances is roughly given (without thermal fluctuations) by the constitutional equation, $\dot{x}^{\mu}=g^{\mu\nu}(x) A_{\nu}(x)$ for $\{\mu, \nu\}=(2, 3)$, which is explicitly written as
\begin{eqnarray}
\begin{cases}
\dot{Y}(t)= -\frac{g_m K}{2N} \left\{ Y^2+Z^2 \right\}^{\sqrt{3}-1} \times Z(t), \\
\dot{Z}(t)= \frac{g_m K}{2N} \left\{Y^2+Z^2 \right\}^{\sqrt{3}-1} \times Y(t),

\end{cases}
\end{eqnarray}
giving an oscillatory motion between two chemical substances like Belousov-Zhabotinsky reaction.  ($Y$ and $Z$ are number of molecules of two chemical solutes in a solvent.)  The angular frequency $\omega$ of this chemical oscillation is given by
\begin{eqnarray}
\omega= \frac{g_m K}{2N} \left\{ Y^2+Z^2 \right\}^{\sqrt{3}-1}.
\end{eqnarray}

Therefore, the monopole like solution seems to be important for the oscillatory chemical reaction in the non-equilibrium thermodynamics, while the black-hole like solution seems to be important for the critical behavior in the thermodynamics.

\section{Fluctuation-dissipation theorem}

The fluctuation-dissipation theorem\cite{fluctuation-dissipation}\cite{Onsager} is well known. Therefore, what we have to do here is only to examine how it works in our gravity analog model.

Following Onsager\cite{Onsager}, we start to estimate
\begin{eqnarray}
\left\langle (x_{\mu}-x_{\mu, 0}) \frac{\delta}{\delta x_{\nu}(t)} \ln \Psi(x, t \vert x_0, -\infty) \right\rangle,  \label{a expectation value}
\end{eqnarray}
where $x_{\mu, 0}$ is an equilibrium value of the thermodynamic variable $x_{\mu}$, and the system is assumed to be in the thermal equilibrium at $t=-\infty$.  Using integration by parts, this expectation value is equal to $\delta_{\mu}^{\nu}=g_{\mu}^{\nu}(x)$.  The expectation value becomes
\begin{eqnarray}
(\ref{a expectation value})=\frac{\int_{-\infty}^{t} \mathcal{D}x^{\mu}(t') ~(x_{\mu}(t)- x_{\mu, 0}(t)) \frac{\delta}{\delta x_{\nu}(t)} e^{-\frac{1}{2k_B T}\int_{-\infty}^{t} \mathcal{L}_{1, L}^{\star}}}{\int_{-\infty}^{t} \mathcal{D}x^{\mu}(t')~e^{-\frac{1}{2k_B T}\int_{-\infty}^{t} \mathcal{L}_{1, L}^{\star}}}.
\end{eqnarray}
We know that
\begin{eqnarray}
&&\frac{\delta}{\delta x_{\nu}(t)} e^{-\frac{1}{2k_B T}\int_{-\infty}^{t} dt' \mathcal{L}_{1, L}^{\star}} \nonumber \\
&&=\frac{\partial}{\partial x_{\nu}(n)} e^{-\frac{1}{2k_B T}\epsilon  \left\{ \frac{1}{2} \left(\frac{ x^{\mu}(n)-x^{\mu}(n-1)}{\epsilon}\right)^2 -  \frac{x^{\mu}(n)-x^{\mu}(n-1)}{\epsilon} A_{\mu}(n-1)+ \frac{1}{2} (A_{\mu}(n-1))^2 + \cdots \right\} } \nonumber \\
&&=-\frac{1}{2k_B T} \left\{ \dot{x}_{\nu} - A_{\nu}(x) \right\} e^{-\frac{1}{2k_B T}\int_{-\infty}^{t} dt' \mathcal{L}_{1, L}^{\star}}
\end{eqnarray}
in the limit of the time interval $\epsilon \to 0$.  Here, we disretize time $t$ as $t=t_n > t_{n-1} > \cdots > -\infty$ with the interval $\epsilon$, and denote such as $A_{\mu}(x(t_i))=A_{\mu}(i)$.  In the above, for $\dot{x}=(x_{i+1}-x_{i})/\epsilon$, $x$ is taken to be $x_i$.  This can be called the ``forward-point prescription''.   A different choice of $x$ is $(x_{i+1}+x_i)/2$ for the same $\dot{x}$, which is usually called the ``mid-point prescription''.  Here, no difference appears between these two descriptions.

Thus we obtain a fluctuation-dissipation theorem of Onsager \cite{Onsager},
\begin{eqnarray}
\langle \left\{x^{\mu}(t)-x^{\mu, 0} \right\} \left\{ \dot{x}^{\nu}(t) - A^{\nu}(x(t)) \right\} \rangle =-2k_BT ~g^{\mu\nu}(x).
\end{eqnarray}
This theorem connects kinetic constants $g^{\mu\nu}(x)$ responsible for the dissipation and the expectation values obtained by summing up the fluctuations.

If we write $\dot{x}$ as a difference, then we have
\begin{eqnarray}
\langle \left\{x^{\mu}(t)-x^{\mu}_{0} \right\} \left\{\left(x^{\nu}(t+\Delta t)-x^{\nu}(t) \right) - \Delta t A^{\nu}(x(t)) \right\} \rangle =-2k_BT \Delta t ~g^{\mu\nu}(x).
\end{eqnarray}

Averaging the above equation over time $t$, by $\lim_{|t''-t'| \to \infty} \frac{1}{t''-t'}\int_{t'}^{t''} dt \cdots $ denoted by overline, we obtain the following result:
\begin{eqnarray}
-2k_BT\Delta t  ~ \overline{g^{\mu\nu}} = \overline{\langle x^{\mu}(t+\Delta t) x^{\nu}(t)\rangle} - \overline{\langle x^{\mu}(t) x^{\nu}(t) \rangle}  -\Delta t \overline{\langle (x^{\mu}(t)-x^{\mu}_{0}) A^{\nu}(x(t)) \rangle },
\end{eqnarray}
where the last term is missing in the previous discussions.

As was imposed by Onsager, if the microscopic reversibility of dynamics holds, then we have
\begin{eqnarray}
\overline{\langle x^{\mu}(t+\Delta t) x^{\nu}(t)\rangle} = \overline{\langle x^{\mu}(t) x^{\nu}(t+\Delta t)\rangle}.
\end{eqnarray}
Furthermore, we have to impose another condition that the overall ``torque'' or ``moment of force" vanishes,
\begin{eqnarray}
\overline{\langle (x^{\mu}(t)-x^{\mu}_{0}) A^{\nu}(x(t)) - (x^{\nu}(t)-x^{\nu}_{0}) A^{\mu}(x(t) \rangle }=0,
\end{eqnarray}
then, the time averaged kinetic constants can be shown symmetric,
\begin{eqnarray}
\overline{g^{\mu\nu}}= \overline{g^{\nu\mu}}.
\end{eqnarray}
In our formulation, the Lagrangian $\mathcal{L}_{1, L}^{\star}$ is written in terms of the symmetric metric, $g_{\mu\nu}(x)$, having the analogy to gravity, but we can add the antisymmetric part of it.  Only the place we can introduce the antisymmetric part $g_{\mu\nu}^{(-)}(x)$ of $g_{\mu\nu}(x)$ is
\begin{eqnarray}
\int ~g_{\mu\nu}^{(-)}(x) (dx^{\mu}  A^{\nu}(x)-dx^{\nu} A^{\mu}(x) )
\end{eqnarray}
which gives the contribution to the ``energy'' by  the ``torque'' of the thermodynamic force $A^{\mu}$.

It is interesting that the above proof which follows faithfully Onsager\cite{Onsager}, can accommodate the anti-symmetric case of kinetic coefficients.  As was discussed in \cite{Onsager}, by taking a monomolecular triangle reaction between three phases $(A, B, C)$ of a chemical substance, the assumption of detailed balance is stronger than to keep the equilibrium. The equilibrium can be fulfilled, even if a circular reaction exists along a circle $(A \to B \to C\to A)$. This comes from the antisymmetric part of the kinetic constants.

Hence, the reciprocal relation in kinetic coefficient is confirmed in our formulation, if the time average of the thermal torque vanishes.

\section{A sketch on how to derive gravity analog thermodynamics from quantum mechanics}
We surely wish to understand how our gravity analog thermodynamics can be derived from quantum mechanics.  This is, however, a very difficult problem.  For this purpose, it is recommended to follow the standard method studied well so far \cite{from QM to SM}.
In the following we give a sketch on this problem based on an analogy to the ``problem of resonance'', since we think it helpful to understand the essence of the problem.  As usual, we will start with the density matrix $\hat{\rho}$, since it is the key ingredient of the problem:
\begin{eqnarray}
\hat{\rho}(t)= e^{-\frac{i}{\hbar} \hat H t } ~\hat{\rho}(0)~ e^{+\frac{i}{\hbar} \hat H t},
\end{eqnarray}
where the Hamiltonian $\hat{H}$ is a microscopic or a quantum mechanical one, and the macroscopic devices such as to generate electric current {\it etc.} are assumed to be included in the system, while the heat bath with temperature $T$ is separated from the system (or from the Hamiltonian).  It is natural to assume that the system is in thermal equilibrium at $t=0$, so that at $t=0$ the density matrix
\begin{eqnarray}
\hat \rho (0)= \sum_{n} \vert n \rangle ~w_n ~\langle n \vert,
\end{eqnarray}
has weights $w_n$ equal to the Boltzmann weights $\omega_n=e^{-E_n/k_B T}$ ($E_n$: the energy eigenvalue of the state $\vert n \rangle$).  We use the wave function $\psi(t)$ to describe the microscopic system by the quantum Hamiltonian $\hat{H}$.  After a certain time $t_{q2c}$ of macroscopic scale has passed (q2c means quantum to classical), we expect the appearance of  the thermodynamics which will be described in terms of the macroscopic thermodynamic variables $x^{\mu}$.

Here, we will remind you of the correspondence between quantum density matrix $\hat{\rho}$ and its classical counterpart, Wigner distribution function $f_W(x, p)$ in  phase space\cite{Wigner}:
\begin{eqnarray}
f_W(x, p; t)=\int_{-\infty}^{\infty} dy ~e^{\frac{i}{\hbar} py} ~\langle x {\scriptstyle + \frac{1}{2}} y \vert \hat{\rho}(t) \vert x {\scriptstyle - \frac{1}{2}} y \rangle. \label{the remained problem}
\end{eqnarray}

The macroscopic variables $\{x^{\mu}\}$ are a collection of generalized coordinates and momenta.  Therefore, we have to separate the even part, $x_{e}=\{x_e^m=x^{2m}\vert m=1, 2, \cdots \}$ from the odd part, $x_{o}=\{x_o^m=x^{2m-1} \vert m=1, 2, \cdots)\}$, where $x^{2m}$ is the momentum canonical conjugate to the coordinate $x^{2m-1}$.

It is successful, if we can show the following equation holds after time $t ~~(\ge t_{q2c})$:
\begin{eqnarray}
f_W(x_o, x_e ; t) \propto \int^{x} \mathcal{D} x^{\mu}(t') e^{-\frac{1}{2k_BT} \int_{0}^{t} dt' \mathcal{L}^{\star}_1}.
\end{eqnarray}

The macroscopic variables $x=(x_o, x_e)$ see the system roughly or macroscopically, that means, during the change of perspective from the microscopic to the macroscopic, some degrees of freedom disappear from the system; the coarse graining occurs or the Hilbert space is reduced.  This situation resembles the problem of resonance or bound state which is formed from the elementary fields. For example, the neutral $K^{0}$ meson is a bound state of a $d$ quark and an anti-particle of $s$ quark, $K^0=(d\bar{s})$, while the antiparticle of $K^{0}$ meson is $\overline{K^0}=(\bar{d}s)$.  The bound states $K^0$ and $\overline{K^0}$, or their mass eigenstates, $K_L$ and $K_S$ decay to quark pairs or to three $\pi$'s $(\pi^{+}\pi^{-}\pi^{0}, \pi^{0}\pi^{0}\pi^{0})$ and two $\pi$'s $(\pi^{+}\pi^{-}, \pi^{0}\pi^{0})$, respectively. If we ignore the microscopic variables of quarks and $\pi$ mesons, the system is described by the macroscopic variables, ($K^0, ~\overline{K^0})$ or $(K_L, ~K_S)$, and the Hamiltonian starts to include the decay widths (rates) of $(K_L, ~K_S)$\cite{neutral K meson}.

Therefore, when the bound states are formed, the Hamiltonian becomes non-hermitian, and the energy becomes complex:
\begin{eqnarray}
E=E_R-\frac{i \hbar}{2}  \Gamma_R,
\end{eqnarray}
where $E_R$ and $\Gamma_R$ are respectively the mass and the decay rate (decay width) of the resonance $R$. This is the Breit-Wigner formula\cite{Breit-Wigner}.

In this way, the reduction of degrees of freedom generates the decay or the dissipation by friction, which occurs equally in the resonance problem as well as in thermodynamics.

Now, we have to estimate
\begin{eqnarray}
&&f_W(x_o, x_e ; t) =\prod_m \int_{-\infty}^{\infty}  dy^m_e ~e^{\frac{i}{\hbar} (x_e y_e)} ~\sum_{n}\langle x_{o} {\scriptstyle+\frac{1}{2}} y_{e} \vert  e^{-\frac{i}{\hbar} \hat H t } ~ \vert n \rangle w_n \langle n \vert~ e^{+\frac{i}{\hbar} \hat H t} \vert x_{o} {\scriptstyle-\frac{1}{2}} y_{e} \rangle ~~~\\
&&=  \prod_m \int_{-\infty}^{\infty} dy^m_e ~e^{\frac{i}{\hbar} (x_e y_e)} ~\sum_{n} w_n \langle n \vert  e^{+\frac{i}{\hbar} \hat H t } ~ \vert x_{o} {\scriptstyle+\frac{1}{2}} y_{e} \rangle^{\dagger} \times \langle n \vert~ e^{+\frac{i}{\hbar} \hat H t} \vert x_{o} {\scriptstyle-\frac{1}{2}} y_{e} \rangle.  \label{classicalization process}
\end{eqnarray}

Using the analogy stated above between the bound state problem and the thermodynamics, the above equation is considered as the product of the decay amplitudes of the resonances; the macroscopic resonances defined by $R_{\pm} \equiv x_{o} {\scriptstyle \pm \frac{1}{2}} y_{e}$ decay to the microscopic state $\psi_n~(=|n\rangle)$.  Therefore, we have to use the Breit-Wigner formula, whose operator form reads
\begin{eqnarray}
\hat{H}=\hat{E}-\frac{i \hbar}{2} \hat{\Gamma},
\end{eqnarray}
where $\hat{E}$ and $\hat{\Gamma}$ are hermitian operators.

The right hand side of Eq.(\ref{classicalization process}) becomes
\begin{eqnarray}
\prod_m \int_{-\infty}^{\infty} dy^m_e ~e^{\frac{i}{\hbar} (x_e y_e)} ~\sum_{n} w_n \langle n \vert e^{+\frac{i}{\hbar} \left(\hat{E}-\frac{i \hbar}{2}\hat{\Gamma}\right) t } ~ \vert R_{+} \rangle^{\dagger} \times \langle n \vert~ e^{+\frac{i}{\hbar} \left(\hat{E}-\frac{i \hbar}{2}\hat{\Gamma}\right)   t} \vert R_{-} \rangle.
\end{eqnarray}
It is not bad to assume here that the energy of the macroscopic resonance, $R_{+}$ and $R_{-}$, and the microscopic states $\psi_n$ to which the resonances decay, have the same energy; that is, $\hat{E}$ is diagonal and $[\hat{E}, \hat{\Gamma}]=0$ holds.  Then, the contribution of the energy levels cancels, remaining only the decay amplitude:
\begin{eqnarray}
\prod_m \int_{-\infty}^{\infty} dy^m_e ~e^{\frac{i}{\hbar} (x_e y_e)} ~\sum_{n} w_n  e^{+\frac{1}{2}\left(\langle n \vert \hat{\Gamma}\vert R_{+} \rangle^{\dagger} + \langle n \vert\hat{\Gamma}\vert R_{-} \rangle \right) t }.
\end{eqnarray}
Here, the decay of $R_{\pm} \to \psi_n$ occurs back in time, which means the generation of macroscopic states occurs along the flow in time from the microscopic states.

If we expand $|R_{\pm} \rangle$ in $y_o$, then we have
\begin{eqnarray}
\prod_m \int_{-\infty}^{\infty} dy^m_e ~e^{\frac{i}{\hbar} (x_e y_e)} ~\sum_{n} w_n  ~e^{t \left\{ \left[1+ \frac{1}{8}\left(y_e \frac{\partial}{\partial x_o}\right)^2 + \cdots \right] Re \Gamma(x_0 \to n) - i \left[\frac{1}{2}\left(y_e \frac{\partial}{\partial x_o} \right) + \cdots \right] Im \Gamma(x_0 \to n) \right\} }.
\end{eqnarray}
Gaussian integration over $y^m_e~(m=1, 2, \cdots)$ yields
\begin{eqnarray}
f_W(x_o, x_e; t) &\propto& \sum_{n} w_n~e^{t \{Re \Gamma(x_o \to n)\}} \nonumber \\
&\times& e^{2t \left\{ \sum_{i, j}\left(\frac{\partial}{\partial x_o^i} Im \Gamma(x_o \to n)\right) \left[ \left(\frac{\partial^2}{\partial x_o^i \partial x_o^j}\right) Re \Gamma(x_o \to n)\right]^{-1} \left(\frac{\partial}{\partial x_o^j} Im \Gamma(x_o \to n)\right) \right\}},  \label{the operator expresion}
\end{eqnarray}
where an approximation of large $t$ is taken. Probably the imaginary part of $\Gamma(x_o \to n)$ is smaller than its real part.  Therefore, we can consider
\begin{eqnarray}
f_W(x_o, x_e; t) \propto \sum_{n} w_ne^{t \left\{ Re \Gamma(x_o \to n) \right\}}=\sum_{n} w_n e^{t \frac{1}{2} \left\{\langle n \vert \hat{\Gamma} \vert x_o \rangle^{\dagger} + \langle n \vert \hat{\Gamma} \vert x_o \rangle \right\}},
\end{eqnarray}
where the sum over $n$ should be taken for the quantum states having the same energy as the macroscopic state; the macroscopic state is written by the thermodynamic variables $x_o=(x^1, x^3, \cdots)$.

An important problem remains, that is the problem to show Eq.(\ref{the remained problem}), by connecting the Winger distribution function $f_W$ to the existence probability of the gravity analog model.  More explicitly, we have to prove
\begin{eqnarray}
\langle x_o \vert e^{t \hat{\Gamma}} \vert x_{o, 0} \rangle \propto
\int_{x_{o, 0}}^{x_o} \mathcal{D} x_o(t') e^{-\frac{1}{2k_BT} \int_{0}^{t} dt' \mathcal{L}^{\star}_1 (x_o, ~\dot{x}_o)}. \label{relation between operator and path integral}
\end{eqnarray}

Although we can not solve this problem completely, we will give some comments in the following.  We first recognize that in the right hand side of the formula Eq.({\ref{the operator expresion}), the operator $\hat{\Gamma}$ has appeared associated with only coordinates $x_o$ (without momentum $x_e$).  It is a good indication, since our gravity analog model is defined using the path integral over the thermodynamic variables ($x_o$).

On the basis of the well known relationship between the operator formalism and the path integral formalism in quantum mechanics, the path integral expression can be obtained from its operator version, by the following trick:
Introduce the thermodynamic Lagrangian $\mathcal{L}^{\star}$ so that it may satisfy $\mathcal{L}^{\star}= (x_e \dot{x}_o)-\Gamma$, and the operator of the decay amplitude $\hat{\Gamma}$ can be obtained from a c-number $\Gamma$ by the replacement given in Appendix, namely
\begin{eqnarray}
\hat{x}_e = -(2k_B T) \frac{\partial}{\partial x_o},
~~\mathrm{and}~~\hat{E}=(2k_BT) \partial_t.
\end{eqnarray}
The role of $\hbar$ in quantum mechanics is played by $(2k_BT)$ in thermodynamics, so that the uncertainty principle (thermal fluctuations) in thermodynamics reads
\begin{eqnarray}
&&\sqrt{\Delta (\hat{x}_o)^2} \sqrt{\Delta (\hat{x}_e)^2} \ge k_B T,  \\
\mathrm{and}&& \sqrt{(\Delta t)^2} \sqrt{(\Delta \hat{E})^2} \ge  k_B T,
\end{eqnarray}
and the correspondence between the operator formalism Hamiltonian ({\it i.e.} decay rate) in thermodynamics and its classical one is related by
\begin{eqnarray}
\Gamma_{\mathrm{classical}} \leftrightarrow \frac{1}{2k_BT} \hat{\Gamma}_{\mathrm{quantum}}.
\end{eqnarray}
In the above correspondence, the direction of the light should be reversed.
Now, the power of the exponential factor in thermodynamics becomes real, and $2k_B T$ appears in the place of $\hbar$ in quantum mechanics as in Eq.(\ref{relation between operator and path integral}).

Finally, we have to determine the form of the gravity analog model.  It takes a quite reasonable form, since its Lagrangian includes functions of $x$ as $A_{\mu}(x)$, as well as $\dot{x}$ and $(\dot{x})^2$, which can be justified, when the thermodynamic variables change slowly in time.

\section{Conclusion and discussions}
We consider in this paper that the space of thermodynamic variables $\{x^{\mu}\}~(\mu=1, 2, \cdots, n)$ forms a manifold which we call Thermodynamic Space (ThS), in which kinetic coefficients (conductivities) play the role of metric tensor $g_{\mu\nu}(x)$.  The Onsager-Machlup and Hashitsume formalism of non-equilibrium thermodynamics\cite{Onsager-Machlup}\cite{Hashitsume} fits quite well to this consideration, in which the currents $\dot{x}^{\mu}$ behave as contravariant vector and the thermodynamic forces $A_{\mu}(x)$ as covariant vector fields. In this way a gravity analog model of non-equilibrium thermodynamics is defined.  The model is easily derived from the Langevin equation following Parisi and Sourlas\cite{Parisi-Sourlas}.

We consider that the metric tensor $g_{\mu\nu}(x)$ and the vector field $A_{\mu}(x)$ also fluctuate thermally in addition to the fluctuation of the path of thermodynamic variables $x^{\mu}(t)$. Then the actions of $g_{\mu\nu}(x)$ and $A_{\mu}(x)$ can be introduced so as to control the fluctuations with a weight $e^{-I^{\star}_2}$.  The strength of the action is described by two constants, the temperature $T$ and the relaxation time $\tau$ (time required to attain the thermal equilibrium).  For example, the strengths of the action of Onsager-Machlup and of Hashitsume, of the typical action of vector field (Maxwell), and of the typical action of gravity (Einstein) for $n=4$ are, respectively, $1/(2k_BT), 1/(k_BT)^2$, and $1/(\tau k_BT)$. This means that if the energy scale $E$ and the time scale $t$ of the problem are equal to $k_BT$ and $\tau$, then three actions contribute equally. If the time scale $t$ is extremely larger than $\tau$, $t >> \tau$, then the gravity is less dominant.  In the usual situation with $t \sim \tau$, the effect of gravity and curved space can not be ignored.  On the contrary if the relevant time scale $t << \tau$, then the gravity contribution dominates.  This supports the analysis of entropy production in Sec.6, in which the gravitational contribution to the entropy production dominates for the rapid operation of the thermodynamic cycle. This is completely different from the weakness of the gravitational effects in nature.

The extension to include non-linear response is easy.  As was discussed in Sec. 5, the dimensional counting is useful to elucidate under what condition the non-linear response dominates.  As an example the non-linear force $A^{\mu} A_{\lambda} A^{\lambda}$ has an extra factor $\tau/k_BT$ relative to the linear response force $A^{\mu}$.  Therefore, if the energy scale $E$ and the time scale $t$ satisfies $E/(k_BT) >> t/\tau$, then the non-linear response dominates; this is the case of slow relaxation time phenomena at low temperature.

 As a candidate of the action $I^{\star}_2$, which gives a weight to the configuration of $g_{\mu\nu}(x)$ and $A_{\mu}(x)$, the effective action induced by the non-equilibrium thermodynamics of Onsager-Machlup and Hashitsume, is derived.  The effective action, obtained in four-dimensional ThS, consists of  $(F_{\mu\nu})^2, (\nabla_{\mu} A_{\nu})^2$ and $R_{\mu\nu} A^{\mu}A^{\nu}$, having a common dependence like $\tau^2/(k_B T)^2$.

We analyze three contributions of the ``entropy production'' caused by a circular thermodynamic process along a circle $C$.  Two terms $\Delta_C \hat{S}_1$ and $\Delta_C \hat{S}_2$ are time reversal even, while the third term $\Delta_C \hat{S}_3$ is time reversal odd.  These three terms have different dependency on the period $P$ required to operate a cyclic process along a circle $C$.
Different $P$ dependency can be used to separate the three entropy productions.  The $P$ independent third term $\Delta_C \hat{S}_3$ has a topological origin and counts the number of lines of magnetic flux passing through the circle, so that the magnetic monopole in ThS is a source of this entropy production.  This entropy production is time reversal odd,  and it can not be cancelled between the forward and the backward operations of the thermodynamic process. The source of the first contribution $\Delta_C \hat{S}_1$, being dominant for the rapid operation $P << 1(=\tau)$, is the gravitational instantons. This entropy production occurs rapidly compared to the relaxation time, so that it can not be supplied by the heat flow. The gravitational instanton is a solution of Euclidean-like metric, since the metric of ThS is positive definite to guarantee the positivity of the kinetic constants.  The second term $\Delta_C \hat{S}_2$ remains even for the quasi-equilibrium process of $P >> 1(=\tau)$, so that it gives the usual increase of the entropy supplied by the heat flow.

 In order to examine the importance of monopole-like and black-hole-like configurations, we give an example in a chemical reaction in a solvent.  For the four-dimensional ThS with variables $(T, V, Y, Z)$, where $(T, V)$ are temperature and volume of a solvent, while $(Y, Z)$ are number of molecules of two chemical substances, the black-hole like metric singularity appears in $(T, V)$ if the solvent follows the van der Waals fluid/gas, while the monopole-like configuration gives an oscillatory reaction between two chemical substances.

The fluctuation-dissipation theorem is examined, for which the Onsager's original argument works quite well, but the standard assumption that the fluctuations are independent of the thermodynamic variables, $\langle x^{\mu} \xi^{\nu} \rangle=0$, does not work well and can not be used.

A sketch to derive the gravity analog model from quantum mechanics is given. The path integral representation of the non-equilibrium thermodynamics is given by the exponential decay with the decay rate (or decay width) $\Gamma(x \to n)$, giving the decay of the thermodynamic variable $x$ into the quantum state labeled by the energy level $n$.  In considering this problem, the operator formalism of thermodynamics is useful, in which $(2k_BT)$ plays the same role as $\hbar$ in quantum mechanics.  This is explained in Appendix, where the Fokker-Planck equation is examined in relation to the operator formalism.

The problems which we do not study well in this paper are listed in the following:

(1) The effect of the Jacobian appeared when replacing the variables from fluctuations $\xi^{\mu}(x, t)$ to the thermodynamic variables $x^{\mu}(t)$ can not be studied well.  It does exist and affords the Parisi-Sourlas supersymmetry\cite{Parisi-Sourlas}.  The Fokker-Planck equation should also be expressed in terms of a fermionic field $\psi(x, t)$ in addition to the bosonic field $\phi(x, t)$ which represents the existence probability of a thermal state.  The system should be super-symmetric under the exchange of  $\phi(x, t)$ and $\psi(x, t)$.  The meaning of the supersymmetry in thermodynamics is not yet clear.

(2) An example of our gravity analog model is given, taking a toy model in a chemical reaction in a solvent, but the detailed analysis of the model (the OUJ model) and others should be done \cite{forthcoming paper}. The toy model suggests that a black-hole like configuration is relevant to the van der Waals liquid/gas, and that a monopole like configuration is relevant to the chemical oscillatory reaction.  It is not clear enough, however, why and how these configurations appear.  To clarify it, we have to study more deeply how the metric singularity in ThS is correlated to the critical behavior of thermodynamics, that is, what type of critical change in the shape of ThS induces what type of phase transition.  Also we have to elucidate what kind of monopole configuration induces what kind of oscillatory chemical reaction. This is surely an interesting issue for us to clarify in the near future.

(3) Contribution to the entropy production from the gravitational instantons are more thoroughly studied.  It is also necessary to elucidate how such gravitational instantons as well as the monopoles are formed in thermodynamics.

(4) The more smooth connection from the quantum mechanics to thermodynamics should be established.  For this purpose we will examine elsewhere, a model in which both quantum and thermal fluctuations exist, and the former disappears afterwards.

(5) In this paper we have not considered thermal forces as gauge field. See \cite{Katagiri} for this point.  Nevertheless, it is clear that the thermodynamic model of Onsager-Machlup and Hashitsume has a simple structure, that is, it is a model of particle $x^{\mu}(t)$ coupled to a gauge field $A_{\mu}(x)$ gauge invariantly.  Important point is, however, there exists a mass term of the gauge field $A_{\mu}(x)A^{\mu}(x)$ and the gauge symmetry is broken accordingly.  An interesting problem is whether this gauge symmetry breaking is done explicitly or spontaneously.  Of course the latter is more interesting, and we surely introduce an additional scalar field $S(x)$ and its vacuum expectation value $\langle S(x) \rangle =v_S$ to break the symmetry.  For this to work $v_S$ should be related to the temperature $T$, such as $v_S=1/2k_BT$.  The equation of motion for $S$ should reproduce properly the change of the local temperature $T(x, t)$.

(6) In Appendix B, we will find that different way of discretization, the mid-point prescription (Stratonovich calculus in stochastic process) or the forward-point prescription (Ito calculus in stochastic process) gives different Fokker-Planck equation, and also notice that there is a delicate problem on how to choose the classical Lagrangian.  The proper classical Lagrangians to reproduce the Fokker-Planck equation of Onsager-Machlup-Hashitusme are
\begin{eqnarray}
&&\mathcal{L}^{' \star}=\mathcal{L}^{\star}_{\mathrm{Stratonovich}}=\mathcal{L}_{\mathrm{Onsager-Machlup}}-\frac{1}{2}A_{\mu}(x)A^{\mu}(x), \\\
&&\mathcal{L}^{''\star}=\mathcal{L}^{\star}_{\mathrm{Ito}}=\mathcal{L}^{\star}_{\mathrm{Onsager-Machlup}}+\frac{1}{2}A_{\mu}(x)A^{\mu}(x).
\end{eqnarray}
We can not deny the possibility that our treatment of classical Lagrangian especially for the mass term of gauge boson, $A_{\mu}(x)A^{\mu}(x)$, is not appropriate.  We are confident on such mass term when the spontaneously breaking of gauge symmetry occurs as in (5), but have not enough experience on the classical introduction of the term by $\int dt A_{\mu}(x(t))A^{\mu}(x(t))$.  If so, how can we remedy this problem?

In field theory it is easy to introduce two scalar fields $\phi(x, t)$ and $S(x, t)$ at the same time.  Can we take a similar way in the classical Lagrangian of point particles $x^{\mu}(t)$? It is easy to introduce two kinds of particles $x^{\mu}(t)$ and $y^{\mu}(t)$ and start with a classical action,
\begin{eqnarray}
I^{\star}&=& \int dt~\frac{1}{2k_BT} \left\{ \frac{1}{2}\dot{x}_{\mu} \dot{x}^{\mu} - \dot{x}^{\mu} A_{\mu}(x, y) + A_{\mu}(x, y)A^{\mu}(x, y) \right\} \nonumber \\
&-&\int dt~\frac{1}{2k_B T'} \left\{ \frac{1}{2}\dot{y}_{\mu} \dot{y}^{\mu} - \dot{y}^{\mu} A_{\mu}(x, y) + A_{\mu}(x, y)A^{\mu}(x, y)\right\},
\end{eqnarray}
where the second action for $y(t)$ have the opposite sign (or have the negative metric) with a different temperature $T'$.

Here, the classical Lagrangian for $x(t)$ and $y(t)$ are chosen as proper ones in the forward-point prescription (See Appendix).  Then, the Fokker-Planck equation is obtained as
\begin{eqnarray}
\partial_t \phi(x, y, t)&=& \left\{ k_BT \frac{\partial}{\partial x_{\mu}}\left(\frac{\partial}{\partial x^{\mu}}-\frac{1}{k_BT} A_{\mu}(x, y) \right) - \frac{1}{2k_BT} (A_{\mu}(x, y))^2
\right\} \phi(x, y, t)  \nonumber \\
& - & \left\{ k_B T' \frac{\partial}{\partial y_{\mu}}\left(\frac{\partial}{\partial y^{\mu}}-\frac{1}{k_BT'} A_{\mu}(x, y) \right) - \frac{1}{2k_BT'} (A_{\mu}(x, y))^2\right\} \phi(x, y, t), ~~~
\end{eqnarray}
in the forward-point prescription (Ito calculus).
 Now we assume that $y^{\mu}(t)$ changes very slowly, so that $y$ can be fixed at a special point and the dependence on $y$ can be ignored.  Then, we have
\begin{eqnarray}
\partial_t \phi(x, t)=\left\{ k_BT~ \partial_{\mu}\left(\partial_{\mu}-\frac{1}{k_BT} A_{\mu}(x) \right) + \left(\frac{1}{2k_BT}-\frac{1}{2k_BT'}\right) (A_{\mu}(x))^2 \right\} \phi(x,t),
\end{eqnarray}

If we choose $T'=\frac{2}{3} T$, then the Fokker-Planck equation in the forward-point prescription (Ito calculus) of the Onsager-Machlup-Hashitsume formalism can be reproduced.  This is a kind of spontaneous breaking mechanism discussed in (5):
\begin{eqnarray}
\langle \phi(x, y, t) \rangle = \langle \phi(x, t) \times S(y, t) \rangle = \phi(x, t) \times v_S.
\end{eqnarray}
This mechanism, however, does not work in the midpoint prescription.

Anyway we will put these problems for future studies.

\section*{Acknowledgements}
The authors give their thanks to Shiro Komata and Ken Yokoyama for a number of valuable comments which help them to clarify the problems involved in this paper. \\
One of the author (AS) thanks Tatsu Takeuchi and Kimiko Yamashita for a fruitful discussion when he rediscovered Parisi-Sourlas supersymmetry several years ago.


\section*{ Appendix A:~Preliminaries for the example }
The example of our gravity analog model is given in the four-dimensional thermodynamic space $(T, V, Y, Z)$ in $M_4$, but it is assumed to be a product of two two-dimensional spaces, one for $(T, V)$ in $M_2$, and the other for $(Y, Z)$ in $M'_2$.
The ansatz for the metric of $M_4=M_2 \times M_2'$ can be
\begin{eqnarray}
ds^2= f(t, r)^2 dt^2 + g(t, r)^2 dr^2 + f'(t', r')^2 dt^{'2} + g'(t', r')^2 dr^{'2},
\end{eqnarray}
where $M_2$ is parametrized by $(t, r)=(T, V)$, while $M'_2$ is parametrized by $(t', r')=(Y, Z)$. $Y$ and $Z$ are the number of molecules of the two chemical substances.
The structure of $M_2$ and $M'_2$ are the same, so that it is enough to know the Christoffel symbols $\Gamma^{\lambda}_{\nu\mu}$ for $M_2$:
\begin{eqnarray}
\Gamma^{0}_{00}= \frac{f_t}{f}, \; \Gamma^{0}_{01}=\Gamma^{0}_{10} =\frac{f_r}{f}, \; \Gamma^{0}_{11}=- \frac{gg_t}{f^2}, \; \Gamma^{1}_{00}= -\frac{f f_r}{g^2},  \; \Gamma^{1}_{01}= \Gamma^{1}_{10}= \frac{g_t}{g}, \; \Gamma^{1}_{11}= \frac{g_r}{g}.
\end{eqnarray}
The $\Gamma$s for $M_2'$ take the same expressions but with $(f, g, t, r)$ replaced by $(f', g', t', r')$.

The Riemann tensor $R^{\mu}_{~\nu\lambda\rho}=\partial_{\lambda}\Gamma^{\mu}_{\nu\rho}-\partial_{\rho}\Gamma^{\mu}_{\nu\lambda}+\Gamma^{\mu}_{\sigma\lambda}\Gamma^{\sigma}_{\nu\rho}-\Gamma^{\mu}_{\sigma\rho}\Gamma^{\sigma}_{\nu\lambda}$ in $M_2$ has an essentially single component $R_{0101}$, so that the Ricci tensor $R_{\mu\nu}=R^{\lambda}_{~\mu\lambda\nu}$ is given by
\begin{eqnarray}
R_{00}=R^1_{~010}=\frac{1}{2} f^2  R=\frac{1}{2} g_{00} R, ~R_{11}=R^0_{~101}=\frac{1}{2} g^2 R=\frac{1}{2} g_{11} R,~\mathrm{and}~R_{01}=R_{10}=0,
\end{eqnarray}
where
\begin{eqnarray}
R=R(t, r)=- \frac{2}{fg} \left\{ \left( \frac{g_t}{f} \right)_t  + \left( \frac{f_r}{g} \right)_r \right\}
\end{eqnarray}
is the scalar curvature, defined by $R= g^{00} R_{00}+ g^{11} R_{11}$.

The Ricci tensor and the scalar curvature for the space $M'_2$ is the same as $M_2$, by replacing  $(f, g, t, r)$ by $(f', g', t', r')$.

Since the covariant derivative is defined by $\nabla_{\mu} A_{\nu}= \partial_{\mu} A_{\nu} - \Gamma^{\lambda}_{\nu\mu} A_{\lambda}$, the symmetric tensor $G_{\mu\nu}$ is given in the space $M_2$ by
\begin{eqnarray}
&&G_{00}= 2\nabla_0 A_0=2 \left\{ (A_0)_t - \frac{f_t}{f} A_0 + \frac{f f_r}{g^2} A_1 \right\}, \\
&& G_{11}=2\nabla_1 A_1= 2 \left\{ (A_1)_r + \frac{gg_t}{f^2} A_0 - \frac{g_r}{g} A_1 \right\}, \\
&&G_{01}=G_{10} =\nabla_0 A_1 + \nabla_1 A_0= (A_0)_r + (A_1)_t - 2 \left\{ \frac{f_r}{f} A_0 + \frac{g_t}{g} A_1 \right\}.
\end{eqnarray}

The field theory Lagrangian is decomposed naturally into the sum of (01) and (23) component Lagrangians, which describe the dynamics in $M_2$ and $M_2'$, respectively,
\begin{eqnarray}
\mathcal{L}^{\star}_2 \propto (\mathrm{Volume~of~} M_2') \times \mathcal{L}^{\star}_{2, \;(01)}  + (\mathrm{Volume~of~} M_2) \times \mathcal{L}^{\star}_{2, \; (23)},
\end{eqnarray}
where
\begin{eqnarray}
&&\mathcal{L}^{\star}_{2,\; (01)} \propto -\frac{1}{6} R(t, r) (A_0 A^0 + A_1 A^1 ) - \frac{1}{8} \sum_{\{\mu\nu\}=0, \; 1} G_{\mu\nu}G^{\mu\nu}, \\
&&\mathcal{L}^{\star}_{2, \; (23)} \propto -\frac{1}{6} R'(t', r') (A_2 A^2 + A_3 A^3 ) - \frac{1}{8} \sum_{\{\mu\nu\}=2, \; 3} G_{\mu\nu}G^{\mu\nu}.
\end{eqnarray}

We begin to examine the first two dimensions, $(T, V)$ of $M_2$, which describes the van der Waals liquid/gas for the solvent with $N$ molecules.
For the space $M_2$, we can impose $f(x^0, x^1) g(x^0, x^1)=1$ or $f(x^0, x^1)=g(x^0, x^1)$.  We will choose the latter $f=g$ which is a familiar choice of describing the two-dimensional world sheet metric in string theory.  Then, we have
\begin{eqnarray}
&&I^{\star}_{2, \; M_2} \propto \int dx^0 dx^1 \sqrt{g} \; \mathcal{L}^{\star}_{2, \; M_2} \nonumber \\
&&\propto \int dx^0 dx^1 \; \frac{1}{3} \left\{ \left(\frac{[f]_{0}}{f}\right) \left[ \frac{1}{f^{2}} \left\{ (A_0)^2 + (A_1)^2 \right\} \right]_0 + \left(\frac{[f]_1}{f} \right)  \left[ \frac{1}{f^{2}} \left\{ (A_0)^2 +(A_1)^2 \right\} \right]_1 \right\} \nonumber \\
&& +\int dx^0 dx^1 \; \frac{1}{8f^{2}} \left[  4\left\{ [A_0]_0 - \left(\frac{[f]_0}{f} A_0 - \frac{[f]_1}{f} A_1 \right) \right\}^2 + 4\left\{ [A_1]_1 + \left(\frac{[f]_0}{f} A_0 - \frac{[f]_1}{f} A_1 \right) \right\}^2 \right. \nonumber \\
&&\left. ~~~~~~~~~~~~~~~~~~~~~~+  2  \left\{ [A_0]_1+ [A_1]_0 - 2 \left( \frac{[f]_1}{f} A_0 + \frac{[f]_0}{f} A_1 \right)\right\}^2 \right],
\end{eqnarray}
where the derivatives with respect to $x^0$ and $x^1$ are denoted by $[~~]_0$ and $[~~]_1$, respectively.

\subsubsection*{A0) Dilute gas limit of the solvent}
In this limit $V=x^1=r \to \infty $, then $S \gg p$, or $|A_0| \gg |A_1|$ we can ignore $A_1=-p$ and $x^0=V$-dependency. In this limit, the equation of motion for $f$ reads
\begin{eqnarray}
\frac{1}{f^2} \left( \frac{S^2}{f^2} f_r \right)_r=0.
\end{eqnarray}
Its solution is
\begin{eqnarray}
&&\frac{1}{f(V)}= \frac{1}{f(Nb)} + \int_{Nb}^{V} dr \; \frac{C}{(Nk_B \ln(r-Nb))^2} = \frac{1}{f(Nb)} + \int_0^{V-Nb} dx \; \frac{C}{(Nk_B \ln(x))^2} \nonumber \\
&&= \frac{1}{f(Nb)} + \frac{C}{(Nk_B)^2} \left[ \mathrm{li}(V-Nb) - \frac{V-Nb}{\ln (V-Nb)} \right].
\end{eqnarray}

Accordingly, we obtain the metric (kinetic coefficients) of the dilute van der Waals gas in the dilute gas limit as
\begin{eqnarray}
g^{00}(V)=g^{11}(V)= \frac{1}{f(V)^2}= \left\{ \frac{1}{f(Nb)} + \frac{C}{(Nk_B)^2} \left[ \mathrm{li}(V-Nb) - \frac{V-Nb}{\ln (V-Nb)} \right] \right\}^2.
\end{eqnarray}

The apparent singularity of li$(V-Nb)$ and $1/\ln(V_Nb)$ at $V-Nb=1(=V_0)$ is not a physical one.  Even if the entropy inherent to volume, $\ln (V-Nb)$, vanishes at a certain volume $V_0$, it can be lifted to be non-zero by the temperature contribution to entropy $c_v \ln (T/T_0)$.  The entropy does not become zero except at $T=0$ (the third law of thermodynamics).   Namely, the metric is thought to be non-singular for the dilute gas and liquid.

\subsubsection*{A1) Close-packing limit of the solvent}
In the close-packing limit we take $V=r \to Nb$, where $p \gg S$.
The effective action in this limit becomes
\begin{eqnarray}
I^{\star}_{2, \; M_2} \propto \int dr \; \left\{ \frac{1}{3} (f_r)^2 \frac{p^2}{f^4} +\frac{1}{2} (p_r)^2 \frac{1}{f^2} -\frac{1}{3} (f_r)(p_r) \frac{1}{f^3} \right\}.
\end{eqnarray}
Its equation of motion reads
\begin{eqnarray}
f (f_{rr})-2 (f_r)^2-2 (f f_r) \frac{1}{V-Nb}=0.
\end{eqnarray}

Then, the metric for the close-packing fluid is obtained:
\begin{eqnarray}
 g^{00}(V)=g^{11}(V) =  \mathrm{const, ~ or}~~ C'(V-Nb)^{6},
 \end{eqnarray}
where $C'$ is a numerical constant.  For the nontrivial solution the metric becomes singular at the close-packing limit of $V \to Nb$.  This is similar to the case of  black-hole.

\subsubsection*{A2) Chemical oscillation between two solutes}

The remaining two-dimensions of $M'_2$ describe the chemical reaction and dissipation between two chemical substances, having $x^2=Y$ and $x^3=Z$ molecules, respectively.

For the two-dimensional space $(Y, Z)$ in $M'_2$ which describes the chemical reactions, the effective action is
\begin{eqnarray}
&&I^{\star}_{2, \; M'_2} \propto  \int dx^2 dx^3 \sqrt{g} \; \mathcal{L}^{\star}_{2, \; M'_2} \nonumber \\
&&\propto \int dx^2 dx^3 \; \frac{1}{3} \left\{ \left(\frac{[f']_{2}}{f'}\right) \left[ \frac{1}{f^{'2}} \left\{ (A_2)^2 + (A_3)^2 \right\} \right]_2 + \left(\frac{[f']_3}{f'} \right)  \left[ \frac{1}{f^{'2}} \left\{ (A_2)^2 +(A_3)^2 \right\} \right]_3 \right\} \nonumber \\
&& +\int dx^2 dx^3 \; \frac{1}{8f^{'2}} \left\{ (G_{22})^2 + (G_{33})^2 +  2 (G_{23})^2 \right\}.
\end{eqnarray}

To obtain the oscillatory behavior, we impose the following ansatz:
\begin{eqnarray}
f'(Y, Z)= F(r), \; A_2(Y, Z)= -\frac{g_m}{2N}\;  Z, \; A_3(Y, Z)=\frac{g_m}{2N}\; Y.
\end{eqnarray}
where $F(r)$ is a function of $r \equiv \sqrt{Y^2+Z^2}$.
Then, we have
\begin{eqnarray}
G_{22}=-G_{33}=2 \left(\frac{g_m}{N}\right) \frac{YZ}{r} \left(\frac{F_r}{F}\right), \; G_{23}=G_{32}=-\left(\frac{g_m}{N}\right) \frac{(Y)^2-(Z)^2}{r} \left(\frac{F_r}{F}\right),
\end{eqnarray}
and
\begin{eqnarray}
(G_{22})^2+(G_{33})^2+2 (G_{23})^2=2 \left(\frac{g_m}{N}\right)^2 r^2 \left(\frac{F_r}{F}\right)^2.
\end{eqnarray}

Therefore, the effective action becomes
\begin{eqnarray}
I^{\star}_{2, \; M'_2} \propto \frac{\pi}{6} \left(\frac{g_m}{N}\right)^2 \int dr \left\{ r^3 \left[(1/F)_r \right]^2-2 r^2 (1/F) (1/F)_r \right\}.
\end{eqnarray}

The equation of motion for $1/F$ is obtained
\begin{eqnarray}
(1/F)_{rr} +  \frac{3}{r} (1/F)_{r} - \frac{2}{r^2} (1/F)=0.
\end{eqnarray}

Its solution reads
\begin{eqnarray}
1/F \propto r^{-1 \pm \sqrt{3}}.
\end{eqnarray}

Therefore, we obtain the most probable metric as follows:
\begin{eqnarray}
g^{22}=g^{33}=(1/F)^2= K \left\{ Y^2+Z^2 \right\}^{-1 \pm \sqrt{3}},
\end{eqnarray}
with a constant $K$.  We have two solutions, but the solution with a positive power of $\sqrt{3}-1$ should be taken, since the conductivity does not seem to diverge when the number of molecules $Y$ and $Z$ tend to zero.

The temporal change of chemical substances is controlled by the constitutional equation, $\dot{x}^{\mu}=g^{\mu\nu}(x) A_{\nu}(x)$ for $\{\mu, \nu\}=(2, 3)$, which are explicitly written as
\begin{eqnarray}
\begin{cases}
\dot{Y}(t)= -\frac{g_m K}{2N} \left\{ Y^2+Z^2 \right\}^{\sqrt{3}-1} \times Z(t), \\
\dot{Z}(t)= \frac{g_m K}{2N} \left\{Y^2+Z^2 \right\}^{\sqrt{3}-1} \times Y(t),

\end{cases}
\end{eqnarray}
giving an oscillatory motion between two chemical substances with the angular frequency
\begin{eqnarray}
\omega= \frac{g_m K}{2N} \left\{ Y^2+Z^2 \right\}^{\sqrt{3}-1}.
\end{eqnarray}

This is a monopole like solution, since it has the non-vanishing magnetic flux, $F_{23}=g_m/N \ne 0$, on the two dimensional thermodynamic space of $(Y, Z)$.

\section*{Appendix B:~~Fokker-Planck equation and \\
the operator formalism of thermodynamics}

In this appendix, we will examine the Fokker-Planck equations in our model, and extract the operator relations for thermodynamic variables.  Afterwards, we will restart from the classical discussion of our thermodynamic model and connect it to the Fokker-Planck equation, based on the operator relations.

Following Feynman\cite{Feynman}, it is easy to derive a differential equation which the existence probability $\phi(s, t)$ satisfies.  For this purpose, we estimate the transition probability function of $\Psi(x, t | x', t')$ having a small time interval $x-x'=\eta$ and $t-t'=\epsilon$ (the minimum time interval), then we have\footnote{A Gaussian integration is used;
$\int d\eta^{\mu} \sqrt{g} ~(\eta^{\mu}\eta^{\nu}) e^{-\eta^2/(4 k_BT \Delta t)}/\int d\eta^{\mu} \sqrt{g}~e^{- \eta^2/(4 k_BT \Delta t)}=2k_BT \Delta t ~g^{\mu\nu}$.}
\begin{eqnarray}
&&\phi(x, t+\epsilon)=\phi(x, t) +\Delta t ~\partial_t \phi(x, t) +\cdots \\
&&\propto \int d\eta ~e^{-\frac{1}{2k_BT} \left( \frac{1}{2\Delta t} \eta_{\mu}\eta^{\mu}-\eta^{\mu}A_{\mu}(x-\eta/2) + \frac{\Delta t}{2} A_{\mu}(x-\eta/2)^2 \right)} \times \left( 1 - \eta^{\mu} \partial_{\mu}  +\frac{1}{2}  \eta^{\mu}  \eta^{\nu}\partial_{\mu}\partial_{\nu} +\cdots \right)\phi(x, t). \nonumber
\end{eqnarray}
Here we choose the ``mid-point prescription" of discretization, that is, $A(x-\eta/2)$.  It is noted that depending on the prescription of discretization, the ordering of operators, such as $\partial_{\mu}$ and $A_{\nu}(x)$ differs.

In the usual field theory, we will not adopt other prescriptions than the mid-point one, since only in this prescription the gauge symmetry becomes manifest and the operator ordering of $\partial_{\mu}$ and $x_{\nu}$ becomes simple and symmetric.  In thermodynamics, however, the forward-point prescription, that is, $\dot{x}=(x(t+\epsilon)-x(t))/\epsilon$ and $x=x(t)$ (not $x=(x(t+1)+x(t))/2$) with $\epsilon >0$ is more familiar.  Therefore, we will compare both prescriptions.

From the above equation, the Fokker-Planck equation in the mid-point prescription, is derived as
\begin{eqnarray}
\partial_t \phi(x, t)=\left\{ k_BT \left(\partial_{\mu}-\frac{1}{2k_BT} A_{\mu}(x) \right)^2 - \frac{1}{2k_BT} (A_{\mu}(x))^2 \right\} \phi(x, t), \label{FP(mid-point)}
\end{eqnarray}
where the covariant derivative appears symmetrically.  From this we can understand the thermodynamics is nothing but a gauge theory, but its symmetry is broken in the presence of the temperature dependent ``mass term'' of the gauge field.

To solve this Fokker-Planck equation, we first factor out a gauge dependent part by the so-called ``Wilson line''
\begin{eqnarray}
&&\phi(x, t) = \varphi_1(x, t) \times \varphi(x, t), \\
&&\mathrm{with}~\varphi_1(x, t)=e^{\frac{1}{2k_BT} \int_{x_0}^{x} dx^{\mu} A_{\mu}(x)},
\end{eqnarray}
then $\varphi(x, t)$ satisfies the following dissipation equation,
\begin{eqnarray}
\partial_t \varphi(x, t) = (k_BT) \partial^2 \varphi(x, t) -\frac{1}{k_BT} \Phi^{(-1)}(A) \varphi(x, t),
\end{eqnarray}
where $\Phi^{(-1)}(A)=\frac{1}{2}A_{\mu}(x)A^{\mu}(x)$ is a variant of Lord Rayleigh's  dissipation function, and is a mass term in the usual gauge theory.

Then, $\varphi(x, t)$ can be obtained and hence, we have $\phi(x, t)$
\begin{eqnarray}
\phi(x, t)=e^{-\frac{1}{k_B T} \int_{t_0}^t \Phi^{(-1)}(A(x(t))) + \frac{1}{2k_BT} \int_{x_0}^{x} dx^{\mu} A_{\mu}(x)}  \frac{1}{(4\pi k_BT(t-t_0))^{n/2}} e^{-\frac{(x-x_0)^2}{4k_B T} } \phi(x_0, t_0),
\end{eqnarray}
where $n$ is the number of thermodynamic variables $x$.

As for the forward-point prescription, we have the following Fokker-Planck equation,
\begin{eqnarray}
\partial_t \phi(x, t) = \left\{ (k_BT) \partial_{\mu} \left( \partial^{\mu}-\frac{1}{k_B T} A^{\mu}(x) \right) -\frac{1}{2k_BT} \Phi^{(-1)}(A) \right\} \phi(x, t), \label{F-P(forward-point)}
\end{eqnarray}
where the derivative and the covariant derivative are mixed up.
Then, by factorizing $\varphi_1(x, t)=e^{\frac{1}{k_BT} \int_{x_0}^{x} dx^{\mu} A_{\mu}(x)}$, we have the diffusion equation also in the forward-point prescription,
\begin{eqnarray}
D_t \varphi(x, t)=\left(\partial_t -A^{\mu}(x)\partial_{\mu} \right)\varphi =(k_BT) \partial^2 \varphi(x, t) -\frac{1}{2k_BT} \Phi^{(-1)}(A) \varphi(x, t),
\end{eqnarray}
where advective term (flow term along with the fluid, $\bm{v}\cdot \bm{\nabla}$ in hydrodynamics) appears and $D_t$ is the Lagrange derivative.

In this case we usually introduce a coordinate $\bar{x}^{\mu}(t)$ which represents the present position of a ship, which starts from the position $x$ at $t=0$ and follows the flow of fluid, namely,
\begin{eqnarray}
\frac{d\bar{x}^{\mu}(t)}{dt}=A^{\mu}(\bar{x})  \label{running variable}.
\end{eqnarray}
Then, we can express $\varphi(x, t)$, by using the solution $\varphi'$ without advective term, as $\varphi(x, t)=\varphi'(\bar{x}(t), t)$, where $x^{\mu}=\bar{x}^{\mu}(t=0)$. It is important to note that this equation Eq.(\ref{running variable}) is identical to the ``classical constitutional equation'' without fluctuations.

Therefore, $\phi(x, t)$ in the forward-point prescription yields,
\begin{eqnarray}
\phi(x, t)=e^{-\frac{1}{2k_B T} \int_{t_0}^t \Phi^{(-1)}(A(\bar{x}(t))) + \frac{1}{k_BT} \int_{x_0}^{x} dx^{\mu} A_{\mu}(x)}  \frac{1}{(4\pi k_BT(t-t_0))^{n/2}} e^{-\frac{(\bar{x}(t) -x_0)^2}{4k_B T} } \phi(x_0, t_0).
\end{eqnarray}

Here, we examine the conservation of probability in the forward-point prescription, or in the usual prescription in thermodynamics.  The current is defined by,
\begin{eqnarray}
j_{\mu}(x, t) =-\left(\partial_{\mu} -\frac{1}{k_BT} A_{\mu}(x) \right) \phi(x, t).
\end{eqnarray}

Then, the conservation law is easily derived from the Fokker-Planck equation:
\begin{eqnarray}
\left(\partial_t + \frac{1}{2k_BT} \Phi^{(-1)}(A) \right) \phi(x, t) + k_BT ~\partial_{\mu} j^{\mu}(x, t)=0.
\end{eqnarray}
Choosing the following combinations for the probability $\phi(x, t)$ and its current $j_{\mu}(x, t)$,
\begin{eqnarray}
&&\phi'(x, t) = e^{-\frac{1}{2k_BT} \int^t \Phi^{(-1)}(A(x(t)))}\phi(x, t), \\
&&j'_{\mu}(x, t)=e^{-\frac{1}{2k_BT} \int^t \Phi^{(-1)}(A(x(t)))}j_{\mu}(x, t),
\end{eqnarray}
they satisfy the simple conservation law:
\begin{eqnarray}
\partial_t \phi'(x, t) +k_BT ~\partial_{\mu} j^{'\mu}(x, t)=0.
\end{eqnarray}

This indicates that if we separate the probability density of the state $\phi(x, t)$ into two factors,
\begin{eqnarray}
\phi(x, t) = e^{-\frac{1}{2k_BT} \int^t dt \Phi^{(-1)}(A(x(t)))}\phi'(x, t),
\end{eqnarray}
then the second part $(\phi'(x, t), j^{'\mu}(x,t))$ reproduces the standard conservation of probability, where the fluid flow with velocity $A^{\mu}(x)$ should be taken into account. Such smooth fluid flow is interrupted, if a ``vortex''($\mathrm{rot}A(x) $) arises, or if the space itself is curved, and the entropy productions $\Delta_C \hat{S}_3$, or $\Delta_C \hat{S}_2$, appears. The contribution of $\Phi^{(-1)}(A(x(t)))$ gives the other entropy production, $\Delta_C \hat{S}_1$, which is a steady production of entropy, occurring even without the thermodynamic flow.

Now, we restart from the classical description of our gravity analog model; its Lagrangian $\mathcal{L}^{\star}_1$ reads


\begin{eqnarray}
\mathcal{L}^{\star}_1(x, \dot{x})=\frac{1}{2} g_{\mu\nu}(x) \dot{x}^{\mu}\dot{x}^{\nu} + \frac{1}{2} g^{\mu\nu}(x) A_{\mu}(x) A_{\nu}(x) -\dot{x^{\mu}} A_{\mu}(x),
\end{eqnarray}
from which we can easily obtain the momentum $p_{\mu}$ as
\begin{eqnarray}
p_{\mu} \equiv \frac{\delta \mathcal{L}^{\star}_1}{\delta \dot{x}^{\mu}}=\dot{x}_{\mu}-A_{\mu}(x)=\xi_{\mu}(x, t).
\end{eqnarray}
Therefore, the microscopic random force $\xi_{\mu}$ is nothing but the momentum.

Hamiltonian $\mathcal{H}$ is defined as usual by
\begin{eqnarray}
\mathcal{H}^{\star}_1= p_{\mu}\dot{x}^{\mu}-\mathcal{L}^{\star}_1,
\end{eqnarray}
from which we obtain
\begin{eqnarray}
&&\mathcal{L}^{\star}_1=\frac{1}{2} p_{\mu}p^{\mu}, ~~\mathrm{and}\\
&&\mathcal{H}^{\star}_1= \frac{1}{2} p_{\mu}p^{\mu}+ \frac{1}{2}(p_{\mu}A^{\mu}(x)+A_{\mu}(x)p^{\mu}) = \frac{1}{2} (p_{\mu}+ A_{\mu}(x))(p^{\mu}+ A^{\mu}(x))-\frac{1}{2}A_{\mu}(x)A^{\mu}(x).~~~~~~~~
\end{eqnarray}
This usual classical expression corresponds to the mid-point prescription. For the time being, we will proceed with this prescription, that is, by expressing $x$ and $p$ ``as symmetric as possible''.

To introduce the operator for $\mathcal{H}^{\star}_1$, the thermal commutation relation should be introduced by
\begin{eqnarray}
[\hat{x}_{\mu}, \hat{p}^{\nu}] = 2 k_B T ~\delta_{\mu}^{\nu}.
\end{eqnarray}
Then, the momentum operator becomes $\hat{p}_{\mu}=-2k_B T \partial _{\mu}$ and we have
\begin{eqnarray}
\frac{1}{2k_BT}\hat{\mathcal{H}}^{\star}_1= k_BT  \left( \partial_{\mu}-\frac{1}{2k_BT} A_{\mu}(x)\right) \left( \partial^{\mu}-\frac{1}{2k_BT} A^{\mu}(x)\right)-\frac{1}{2k_BT}\times \frac{1}{2}A_{\mu}(x)A^{\mu}(x).
\end{eqnarray}
This does not reproduce the right-hand-side of the Fokker-Planck equation in the mid-point prescription.
This discrepancy should be remedied in the classical Lagrangian, and we find that the correct choice of classical Lagrangian in the mid-point prescription is
\begin{eqnarray}
\mathcal{L}^{\star}_1(x, \dot{x})'=\frac{1}{2} g_{\mu\nu}(x) \dot{x}^{\mu}\dot{x}^{\nu} -\dot{x^{\mu}} A_{\mu}(x).
\end{eqnarray}
The Hamiltonian for this Lagrangian reads,
\begin{eqnarray}
\mathcal{H}^{'\star}_1= \frac{1}{2} p_{\mu}p^{\mu}+ \frac{1}{2}(p_{\mu}A^{\mu}(x)+A_{\mu}(x)p^{\mu}) = \frac{1}{2} (p_{\mu}+ A_{\mu}(x))(p^{\mu}+ A^{\mu}(x))-A_{\mu}(x)A^{\mu}(x).
\end{eqnarray}

The corresponding quantum Hamiltonian is
\begin{eqnarray}
\frac{1}{2k_BT}\hat{\mathcal{H}}^{'\star}_1= k_BT  \left( \partial_{\mu}-\frac{1}{2k_BT} A_{\mu}(x)\right) \left( \partial^{\mu}-\frac{1}{2k_BT} A^{\mu}(x)\right)-\frac{1}{2k_BT}A_{\mu}(x)A^{\mu}(x)
\end{eqnarray}
which coincides with the r.h.s. of the Fokker-Planck equation, and we have
\begin{eqnarray}
2k_BT ~\partial_t \phi(x, t)= \hat{\mathcal{H}}^{'\star}_1 \phi(x, t).
\end{eqnarray}
Now, the operator relation in the operator formalism of thermodynamics is obtained in the mid-point prescription as
\begin{eqnarray}
\hat{p}_{\mu}= -2k_B T \partial_{\mu},~\mathrm{and}~\hat{E}_{\mu}= 2k_B T \partial_{t}
\end{eqnarray}
which guarantees the commutation relation and uncertainty principle stated in Sec. 8.

One more thing we have to check is the forward-time prescription used popularly in thermodynamics.
In this case the correct classical Lagrangian is
\begin{eqnarray}
\mathcal{L}^{\star}_1(x, \dot{x})''=\frac{1}{2} g_{\mu\nu}(x) \dot{x}^{\mu}\dot{x}^{\nu} -\dot{x^{\mu}} A_{\mu}(x)+A_{\mu}(x) A^{\mu}(x),
\end{eqnarray}
the momentum is the same, and the Hamiltonian is
\begin{eqnarray}
\mathcal{H}^{''\star}_1= \frac{1}{2} p_{\mu}(p^{\mu}+ 2A_{\mu}(x))-\frac{1}{2}A_{\mu} A^{\mu}.
\end{eqnarray}
Then, the r.h.s of the Fokker-Planck equation in the forward-point prescription (or the standard prescription in thermodynamics) becomes
\begin{eqnarray}
\frac{1}{2k_BT}\hat{\mathcal{H}}^{''\star}_1= k_BT  \partial_{\mu} \left( \partial^{\mu}-\frac{1}{k_BT} A^{\mu}(x)\right)-\frac{1}{4k_BT}A_{\mu}(x)A^{\mu}(x).
\end{eqnarray}
Thus the Fokker-Planck equation coincides with that obtained by the  operator formalism:
\begin{eqnarray}
2k_BT ~\partial_t \phi(x, t)= \hat{\mathcal{H}}^{''\star}_1 \phi(x, t).
\end{eqnarray}

It is clear in both prescriptions that the operator relations are the same, so that the Onsager-Machlup path integral formalism and the operator formalism (of imposing the commutation relations between thermodynamic variables and their momenta) are consistent.



\begin{thebibliography}{99}

\bibitem{Onsager-Machlup}
L. Onsager and S. Machlup, Phys. Rev. {\bf 91} (1953) 1505.

\bibitem{Hashitsume}
N. Hashitsume, Prog. Theor. Phys. {\bf 8} (1952) 461; {\it ibid.} {\bf 15} (1956) 369; ``Proc. Int. Conf. Theor. Phys. Kyoto Sept.'' (1953) p.495; An essay ``Four ways of describing materials and two kinds of entropy'' (in Japanese) in ``Development of the second law of thermodynamics'' ed. by S. Ono, A. Tsuchida, T. Murota and E. Yagi, Asakura Publishing Company (1990).

\bibitem{Parisi-Sourlas}
G. Parisi and N. Sourlas, Phys. Rev. Lett. {\bf 43}, (1979) 744.

\bibitem{BRST}
C. Becchi, A. Rouet and R. Stora, Ann. Phys. {\bf 98} (1976) 2; \\
I. V. Tyutin, Lebedev Physics Institute preprint No. 39 (1975), arXiv:0812.0580.

\bibitem{Baulieu}
L. Baulieu, Prog. Theor. Phys. Supp. {\bf 111} (1993) 151.

\bibitem{detailed fluctuation theorem}
D. J. Evans, E. G. D. Cohen and G. P. Morriss, Phys. Rev. Lett. {\bf 71} (1993) 2041; \\
G. E. Crooks, Phys. Rev. {\bf E 60} (1999) 2721.

\bibitem{Brevin}
See a good explanation, L. Brevin, ``Riemann Normal Coordinates'' (1996) [http://users.monash.edu.au/~leo/research/papers/files/lcb96-01.pdf].

\bibitem{gravitational instanton}
G. W. Gibbons and S. W. Hawking, Comm. Math. Phys. {\bf 66} (1979) 291;\\
T. Eguchi, P. G. Gilkey and A. J. Hanson, Phys. Reports {\bf 66} (1986).

\bibitem{fluctuation-dissipation}
As a comprehensive review on the fluctuation dissipation theorem, see R. Kubo, Rep. Prog. Phys. {\bf 29} (1966) 255.

\bibitem{textbook of thermodynamics}
See for example,  Section 5, N. Hashitsume,``Introduction to Thermodynamics and Statistical Mechanics'' (in Japanese) Iwanami Publishing Company (1981).

\bibitem{Ruppeimer}
G. Ruppeimer, Rev. Mod. Phys. {\bf 67} (1995) 605; Erratum {\it ibid.} {\bf 68} (1996) 313.

\bibitem{Sonnino}
G. Sonnino and A. Sonnino, J. Thermodyn. Catal. {\bf 5:2} (2014) 1000129.

\bibitem{Katagiri}
S. Katagiri, PTEP, 093A02 (2018) [arXiv:1806.07816].

\bibitem{proper time formalism}
J. Schwinger, Phys. Rev. {\bf 82}, 664 (1951); \\
V. Fock, Physik. Z. Sowjetunion, {\bf 12}, 404 (1937); \\
Y. Nambu, Prog. Theor. Phys. {\bf 5}, 82 (1950); \\
K. Yamashita, X. Fan, S. Kamioka, S. Asai, and A. Sugamoto, PTEP, 123B03 (2017) [arXiv:1707.03308].

\bibitem{Ignatiev}
A. Y. Ignatiev and G. J. Girish, Mod. Phys. Lett. {\bf{A11} } (1996) 2735.

\bibitem{Onsager}
L. Onsager, Phys. Rev. {\bf 37} (1931) 405; {\it ibid.} {\bf 38} (1931) 2265.

\bibitem{from QM to SM}
J. von Neumann, Zeitschrift f\"ur Physik {\bf 57} (1929) 30 [arXiv:1003.2133];\\
P. C. Martin and J. Schwinger, Phys. Rev. {\bf 115} (1959) 1342; \\
R. P. Feynman and F. L. Vernon, Ann. of Phys. {\bf 24} (1963) 118;\\
L. V. Keldysh, J. Exp. Theor. Phys. {\bf 47} (1964) 1515.

\bibitem{Wigner}
E. P. Wigner, Phys. Rev. {\bf 40} (1932) 749.

\bibitem{neutral K meson}
See for example, L.B. Okun, Chapter 11, ``Leptons and Quarks'' (1984); \\
B. Winstein and L. Wolfenstein, Rev. Mod. Phys. {\bf 65} (1993) 1113; \\
a brief description is found in R. P. Feynman, Chapter 11-5, ``Lectures on Physics-Quantum Mechanics'' Addison-Wesley Publishing Company (1966) .

\bibitem{Breit-Wigner}
G. Breit and E. P. Wigner, Phys. Rev. {\bf 49} (1936) 519.

\bibitem{Feynman}
R. P. Feynman, Rev. Mod. Phys. {\bf 20} (1948) 367;\\
See ``Feynman's Thesis-A New Approach to Quantum Theory'', World Scientific Publishing (2005), which includes Feynman's thesis (1942), the aforementioned paper, and the Dirac's relevant paper (1933), [http://files.untiredwithloving.org/thesis.pdf].


\bibitem{forthcoming paper}
 Noriaki Aibara {\it et al.} (The OUJ Tokyo Bunkyo Field Theory Collaboration), ``A gravity analog model of chemical reaction--OUJ model--'' (in preparation).


\end{thebibliography}
\end{document}